\documentclass[twocolumn,3p,times,procedia,compress]{elsarticle}

\usepackage{ecrc}
\usepackage{graphicx}
\usepackage{balance}
\usepackage{array,threeparttable}
\usepackage{url}

\volume{00}

\firstpage{1}

\runauth{}

\jid{procs}

\usepackage{amssymb}
\usepackage{amsmath}
\usepackage{multicol}
\usepackage{algorithm}
\usepackage{algpseudocode}

\usepackage{subfigure}
\usepackage{ntheorem}
\newtheorem{theorem}{Theorem}
\newtheorem{proposition}{Proposition}
\newtheorem*{Proof}{Proof}

\usepackage[figuresright]{rotating}
\usepackage{bm}
\usepackage{caption}
\usepackage{titlesec}
\titleformat*{\section}{\small \bf}
\titleformat*{\subsection}{\small \em}
\titleformat*{\subsubsection}{\small \em}

\usepackage{fancyhdr}
\fancyheadoffset[RO,EL]{0pt}
\usepackage{geometry}
\begin{document}\small
\begin{frontmatter}

\dochead{}

\title{
\begin{flushleft}
{\LARGE Double-Side  Delay Alignment Modulation for Multi-User Millimeter Wave and TeraHertz  Communications} 
\end{flushleft}
}

\author[]{ \leftline {Xingwei Wang$^a$, Haiquan Lu$^a$$^b$, Jieni Zhang$^a$, Yong Zeng $^*$$^a$$^b$}}

\address{ \leftline {$^a$National Mobile Communications Research Laboratory, Southeast University, Nanjing 210096, China}

  \leftline {$^b$Purple Mountain Laboratories, Nanjing 211111, China}

}

\cortext[]{This work was supported by the Natural Science Foundation for Distinguished Young Scholars of Jiangsu Province with grant number BK20240070, and also by the National  Natural Science Foundation of China under Grant 62071114. (Corresponding author: Yong Zeng.)}


\begin{abstract}

Delay alignment modulation (DAM) is an innovative broadband modulation technique well suited for millimeter wave (mmWave) and terahertz (THz) massive multiple-input multiple-output (MIMO) communication  systems. Leveraging the high spatial resolution and sparsity of multi-path channels, DAM mitigates inter-symbol interference (ISI) effectively, by aligning all multi-path components through a combination of  delay pre/post-compensation and path-based beamforming. As such, ISI is eliminated while preserving multi-path power gains. In this paper, we explore multi-user double-side  DAM with both delay pre-compensation at the transmitter  and post-compensation at the receiver, contrasting with prior one-side DAM that primarily focuses  on delay pre-compensation only. Firstly,  we reveal  the constraint for the introduced delays and the delay pre/post-compensation vectors tailored for multi-user double-side DAM, given a specific number of delay pre/post-compensations. Furthermore, we show that as long as the number of base station (BS)/user equipment (UE) antennas is sufficiently large, single-side DAM, where delay compensation is only performed at the BS/UE, is preferred than double-side DAM since the former results in less ISI to be spatially eliminated. Next, we propose two low-complexity  path-based beamforming strategies based on the eigen-beamforming transmission and ISI-zero forcing (ZF) principles, respectively, based on which the achievable sum rates are studied. Simulation results verify that  with sufficiently large BS/UE antennas, single-side DAM is sufficient. Furthermore, compared to the benchmark scheme of   orthogonal frequency division multiplexing (OFDM), multi-user BS-side DAM achieves higher spectral efficiency and/or lower peak-to-average power ratio (PAPR).

\end{abstract}


\end{frontmatter}

\section{Introduction}

To support the ambitious goals  such as ultra-high data throughput to accommodate the rapid growth in data traffic, as well as new capabilities of integrated sensing and communication (ISAC) \cite{IMT}, various technologies   have been investigated for the sixth generation (6G) mobile communication  networks, such as  new spectrum utilization and new multiple-input multiple-output (MIMO) evolution \cite{CKM}. In particular, high-frequency bands, including millimeter-wave (mmWave) and terahertz (THz) frequencies, exhibit multi-path sparsity, which can be utilized for  achieving ultra-high data rates and positioning/sensing accuracy \cite{MMTHz-1,MMTHz-2,THz-1}. Another potential  technology for 6G is extremely large-scale multiple-input multiple-output (XL-MIMO) \cite{XL-MIMO-1,XL-MIMO-2,XL-MIMO-3} or extremely large-scale aperture arrays (ELAA) \cite{spatial}. These technologies, with array size significantly  larger than that in massive MIMO, may significantly enhance spectral efficiency and provide super spatial resolution, enabling more accurate differentiation of multi-path signals from various directions. Besides, integrated localization, sensing, and communication (ILSAC) is expected to play a crucial role in 6G systems \cite{ISAC-1,ISAC-6}. By synergistically exploiting the sparsity of high-frequency channels and the super spatial resolution of XL-MIMO, ILSAC enables more accurate estimation of  position and orientation of mobile terminals, even the detailed feature extraction for delay/angular information in each multi-path channel \cite{ISAC-1,ISAC-6,ISAC-2,ISAC-3,ISAC-5}.

Note that multi-path propagation is an inherent feature of wireless channel, due to the signal reflection, refraction, scattering or diffraction from obstructions along the path of transmission. To address the inter-symbol interference (ISI) issue caused by multi-path propagation, the different generation mobile communication systems have developed  different approaches. For example, in 2G networks, primarily represented by global system for mobile communications (GSM), ISI mitigation was achieved through equalization techniques, such as time-domain equalizers \cite{equalization}.  However, the complexity of time-domain equalization   increases significantly  with channel delay spread. The introduction of the third generation (3G) networks, particularly  wideband code division multiple access (WCDMA), used  RAKE receivers to leverage multi-path diversity by combining signals from different paths. However, they require spread spectrum codes with good auto/cross-correlation properties and bandwidth expansion, which demands a bandwidth much larger than necessary for data transmission \cite{goldsmith}.  Moreover, the small number of pseudo-orthogonal codes limits the capacity of the 3G networks \cite{2G}.

The fourth generation (4G) and   fifth generation (5G)  networks mainly adopted  orthogonal frequency division multiplexing (OFDM)  for ISI mitigation \cite{goldsmith}. OFDM provides robustness against multi-path interference through the use of orthogonal sub-carriers and cyclic prefixes (CPs). Frequency-domain equalization is inherently facilitated in OFDM due to the parallel transmission of symbols in the frequency domain. Although this simplifies equalization, OFDM is plagued by a high peak-to-average power ratio (PAPR), leading to power inefficiencies, and is sensitive to carrier frequency offset (CFO) as well as  the severe out-of-band (OOB) emission.  To mitigate these challenges, single-carrier frequency-domain equalization (SC-FDE) has been introduced to reduce PAPR, while filter bank multi-carrier (FBMC) aims to minimize OOB emissions \cite{SCC}. However, SC-FDE introduces increased receiver complexity due to the requirement for the fast Fourier transform (FFT) and inverse FFT processing, while FBMC exhibits higher PAPR and complexity resulting from sub-carrier filtering.   Furthermore, the recently proposed    orthogonal time frequency space (OTFS) modulation introduces a new approach by spreading data symbols across time and frequency domains, enhancing robustness against channel variations and Doppler effects. This addresses the severe CFO issue encountered by OFDM in high mobility scenarios \cite{OTFS-1, OTFSbook, OTFS-overview}.

For systems with  large antenna arrays \cite{XL-MIMO-3, spatial} and operating at high-frequency bands such as mmWave \cite{mmWave} and THz  frequencies \cite{SCC}, delay alignment modulation (DAM) was recently proposed as an effective technique for ISI mitigation  \cite{DAM,DAM-OFDM}. Existing literature predominantly centers around single-user DAM scenarios with delay pre-compensations \cite{DAM,DAM-OFDM, DAM-CSI,DAM-ISAC,DAM-ISAC-other,DAM-ISAC-1,DAM-ISAC-2,DAM-IRSs,Hybrid-DAM}. For instance, the work  \cite{DAM} first introduced the concept of DAM, laying the foundation for subsequent investigations. Building upon this framework, \cite{DAM-CSI} proposed a block orthogonal matching pursuit (BOMP)-based channel estimation method tailored for DAM. Moreover, DAM has found applications in ISAC \cite{DAM-ISAC,DAM-ISAC-other,DAM-ISAC-1,DAM-ISAC-2} and communication aided by multi-intelligent reflecting surfaces (IRSs) \cite{DAM-IRSs}. Additionally, \cite{Hybrid-DAM} extended DAM based on  digital beamforming schemes to that based on  hybrid beamforming. In contrast to these studies,  \cite{DAM-OFDM} introduced a unified framework termed DAM-OFDM capable of achieving ISI-free communication using either single- or multi-carrier transmissions. Subsequently, \cite{fDAM} extended DAM by considering  fractional channel delays. Further advancements were made by \cite{Multi-UserDAM, MU-DAM}, which considered DAM for multi-user communication scenarios. 
Compared to OFDM and OTFS, single-carrier DAM is able to achieve lower PAPR and save the CP overheard. Besides, single-carrier DAM enables the instant symbol-by-symbol signal detection at the receiver \cite{DAM-IRSs,DDAM-1}. 	

However, existing works on DAM \cite{DAM,DAM-CSI,DAM-OFDM,DAM-ISAC,DAM-ISAC-other,DAM-ISAC-1,DAM-ISAC-2,DAM-IRSs,Hybrid-DAM,fDAM,Multi-UserDAM, MU-DAM} have primarily concentrated on base station (BS)-side delay compensation. Note that for time or frequency-domain equalziations, employing both  pre-equalization at the transmitter and post-equalization at the receiver may outperform the single-side  equalization at the transmitter or receiver alone \cite{pre-post,pre-post1,pre-post2}. Similar to MIMO communications with multiple antennas at both BS and user equipment (UE), the delay pre-compensation at the transmitter and the delay post-compensation at the receiver can be simultaneously applied, and \cite{TDL} verified  this result.  However, the work \cite{TDL} did not exploit the
high spatial resolution or multi-path sparsity of mmWave/THz massive MIMO systems to eliminate ISI.

\begin{figure}
	\centerline
	{\includegraphics[scale=0.45]{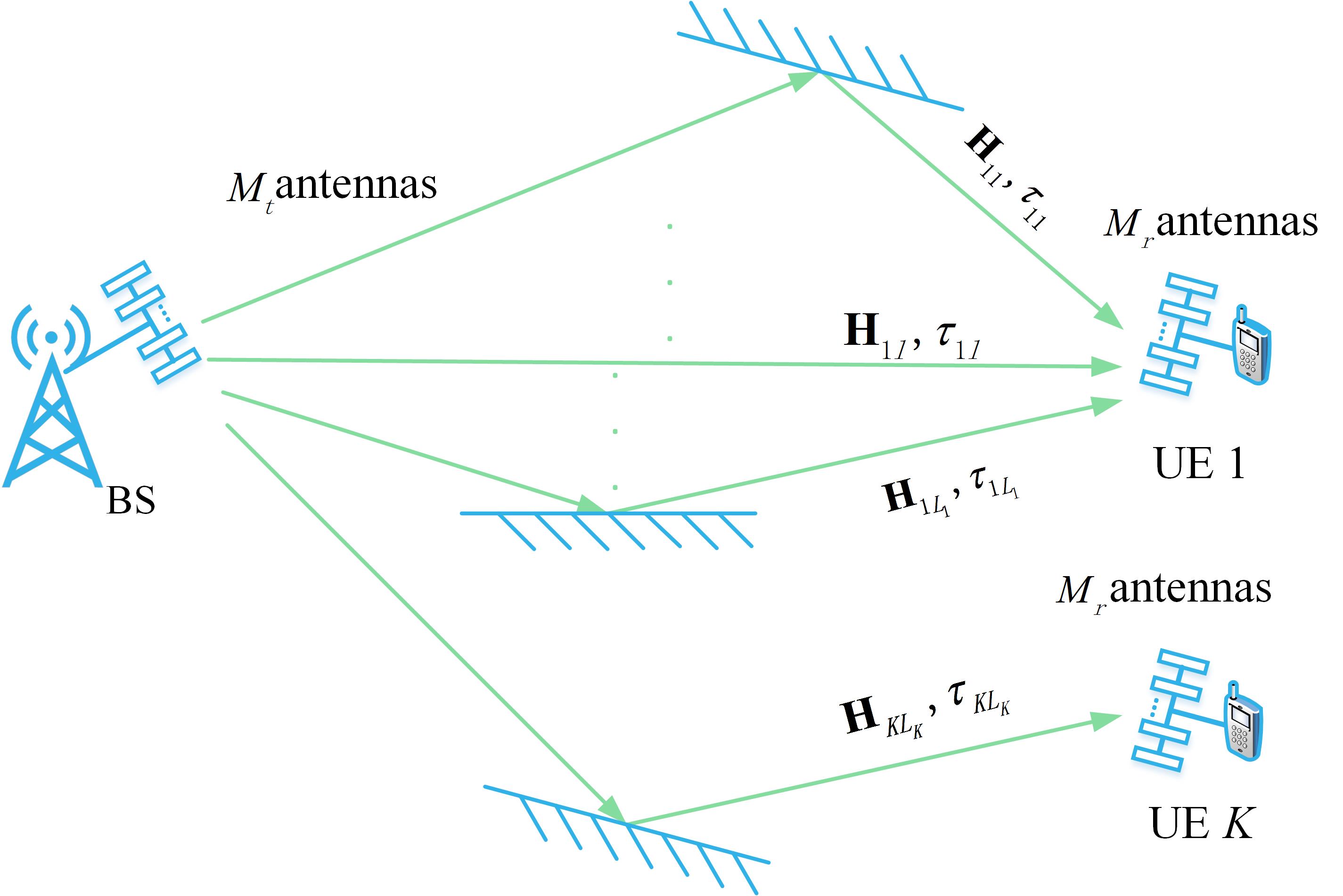}} 
	\caption{A muti-user mmWave/THz massive  MIMO communication system.} \label{fig1}
\end{figure}

In this paper, we aim to bridge the gap by introducing both delay  pre-compensation and post-compensation for multi-user  DAM, termed double-side DAM. The main contributions of this paper are summarized as follows:
\begin{itemize}
	
	\item  First, we establish the input-output relationships for double-side DAM for fractional channel delays in multi-user  communication systems, where   delay pre-compensation  and post-compensation are applied at the transmitter and the  receiver, respectively.

	\item  Second,  we reveal  the constraint for the introduced delays and the pre/post compensation vectors  for multi-user double-side DAM, given a specific number of delay pre/post-compensation. Specifically, we derive  non-homogeneous equations regarding the number of pre-compensation, post-compensation and channel delays, and then  the sufficient condition for the solvability of these equations  of each UE  is determined.  We identify a fixed non-homogeneous equation for simplicity, which allows us to  compute the vectors for both delay pre-compensation and post-compensation. Based on above, we introduce an interference minimization model for each UE. The results indicate that  as long as the number of  BS/UE antennas is sufficiently large, single-side DAM, where delay compensation is only performed at the BS/UE, is preferred than double-side DAM since the former results in less ISI to be spatially eliminated.

	\item 	Third, we propose two low-complexity path-based beamforming strategies: one based on eigen-beamforming transmission and the other based on ISI-zero forcing (ZF) principles. The achievable sum rates for these strategies are  analyzed. Specifically, we first develop a low-complexity path-based eigen-beamforming approach for multi-user DAM that tolerates both ISI and inter-user interference (IUI). Next, we consider the ISI-ZF beamforming strategy, which partially eliminates ISI and completely removes IUI. This approach utilizes an alternating optimization method to maximize the sum rate, starting  with a receiver design based on the minimum mean squared error (MMSE) criterion, followed by   ISI-ZF transmit beamforming.

	\item Last, we compare the proposed multi-user single-carrier DAM designs with the benchmarking of the prevalent OFDM scheme. Extensive simulation results demonstrate the advantages of single-side DAM when the BS antenna is  sufficiently large, in terms of spectral efficiency and/or PAPR.

\end{itemize}

The remainder of this paper is organized as follows. Section 2 presents the system model. Section 3 provides a comprehensive design of delay pre-compensation and post-compensation for multi-user DAM. Section 4 investigates the sum rate performance of BS-side multi-user DAM, comparing it with benchmarking scheme of OFDM. Section 5 presents detailed numerical results. Finally, the conclusions are drawn in Section 6.

 {\it Notations}: Scalars are denoted by italic letters. Vectors and matrices are denoted by boldface lower- and upper-case letters, respectively. For a matrix $\boldsymbol{\mathrm{A}} \in \mathbb{C}^{M \times N}$, $\boldsymbol{\mathrm{A}}^T, \boldsymbol{\mathrm{A}}^H, \boldsymbol{\mathrm{A}}^{\dagger}$, $\|\boldsymbol{\mathrm{A}}\|_F$, $\text{rank}(\boldsymbol{\mathrm{A}})$, and $[\boldsymbol{\mathrm{A}}]_{m,i:j}$ denote its transpose, Hermitian transpose, pseudo-inverse, Frobenius norm, rank, and the elements in the $m$-th row and the columns from $i$ to $j$,  respectively.  For a vector $\boldsymbol{\mathrm{a}} \in \mathbb{C}^{M \times 1}$, $\|\boldsymbol{\mathrm{a}}\|$ and $[\boldsymbol{\mathrm{a}}]_{i:j}$ denote its  $l_2$-norm and the rows from $i$ to $j$, respectively.
$\mathbb{E}[\cdot]$, $\delta(\cdot)$, $*$ and $\lfloor \cdot \rceil$ denote the expectation,  Dirac-delta impulse function, linear convolution, and round to the nearest integer, respectively. $\mathcal{CN}(\boldsymbol{\mathrm{x}},\bf{\Sigma})$  denotes the  circularly symmetric complex Gaussian (CSCG)  distribution of a random vector with mean vector $\boldsymbol{\mathrm{x}}$ and covariance matrix $\bf{\Sigma}$.   $\boldsymbol{\mathrm{0}}_{M \times N}, \boldsymbol{\mathrm{1}}_{M \times N}$, and $\boldsymbol{\mathrm{I}}_{M \times N}$ denote the all-zeros  matrix, the all-ones matrix, and the identity  matrix,  respectively. 
\begin{figure*}
	\centering
	\includegraphics[scale=.5]{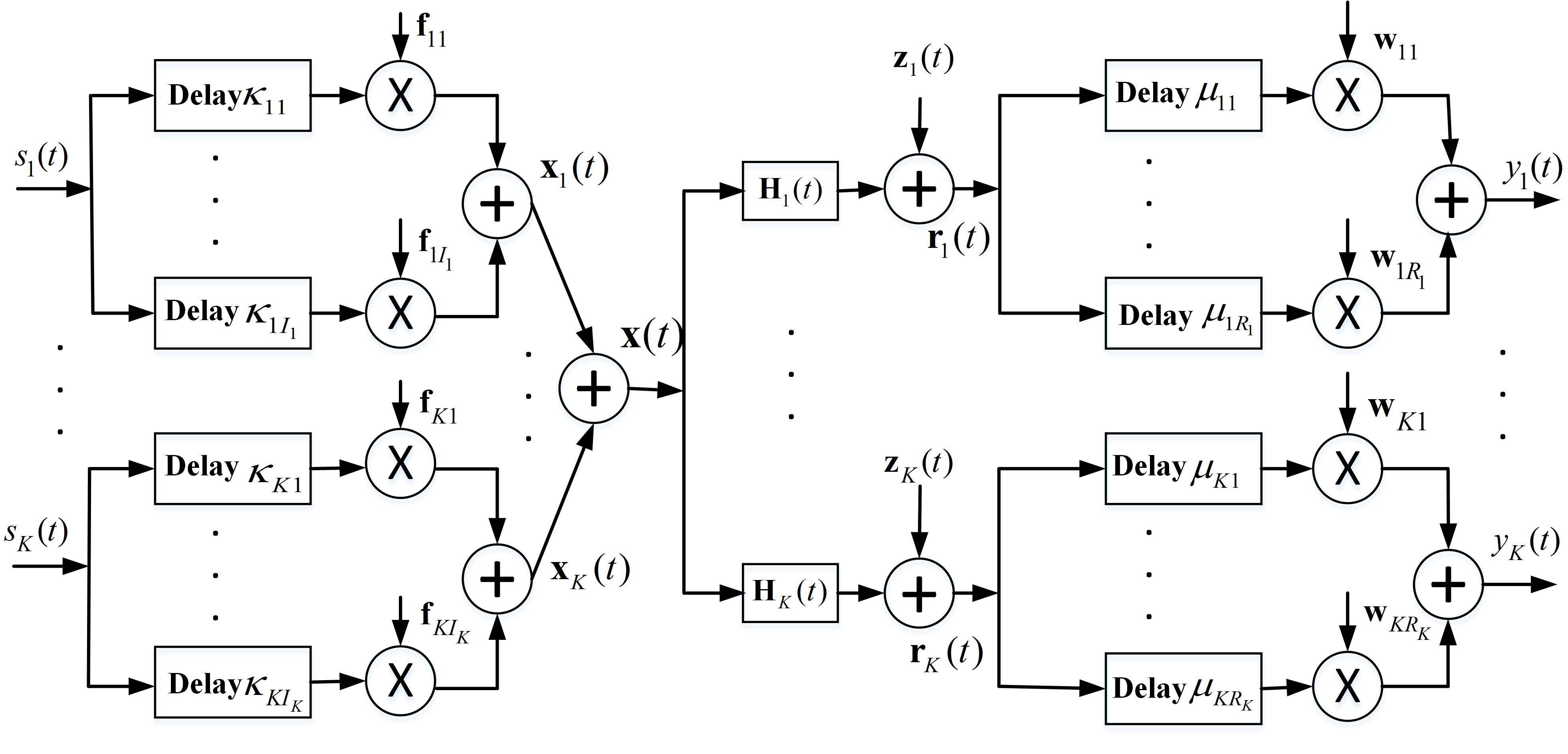}
	\caption{Block diagram for multi-user single-carrier double-side DAM.}\label{fig2}
\end{figure*}
\section{System model} 

As illustrated in Fig. \ref{fig1}, we consider a multi-user mmWave/THz massive MIMO communication system, where the BS is equipped with $M_t$ antennas  serving  $K$ UEs, each with $M_r$ antennas. For multi-path environment, the baseband channel impulse response of UE $k$ within each channel coherence block can be expressed as 
\begin{equation}
	\boldsymbol{\mathrm{H}}_k(t)=\sum_{l=1}^{L_k} \boldsymbol{\mathrm{H}}_{kl} \delta(t-\tau_{kl}), \label{1}
\end{equation}
where  $\tau_{kl}=n_{kl}T+{\tau}_{\text{F},kl}$, and $T$ is the signal sampling interval, with $n_{kl}=\lfloor \frac{{\tau}_{kl}}{T}\rceil$ and ${\tau}_{\text{F},kl}=({\tau}_{kl}-n_{kl}T)\in [-\frac{T}{2},\frac{T}{2}]$ corresponding to the integer and fractional parts of the delays, respectively; $\boldsymbol{\mathrm{H}}_{kl} \in \mathbb{C}^{M_r \times M_t}$ denotes the channel matrix of the $l$-th multi-path for UE $k$, and $L_k$ is the number of temporal-resolvable multi-paths of UE $k$. Without loss of generality, we assume that $n_{k1}<n_{k2}<,...,<n_{kL_k}$.

Let $s_k[n]$ be the independent and identically distributed (i.i.d.) information-bearing symbols with normalized power $\mathbb{E}[|s_k[n]|^2]=1$. The general block diagram  for  single-carrier double-side  DAM is illustrated  in Fig. \ref{fig2}. The  transmitted discrete-time signal is \cite{Multi-UserDAM}
\begin{equation}
	\boldsymbol{\mathrm{x}}[n]=\sum_{k=1}^K\sum_{i=1}^{I_k}\boldsymbol{\mathrm{f}}_{k{i}} s_k[n-\kappa_{k{i}}], \label{2}
\end{equation}
where $I_k$ denotes the number of pre-compensation delays for UE $k$, $\boldsymbol{\mathrm{f}}_{ki} \in \mathbb{C}^{M_t \times 1}$ denotes the  path-based beamforming   vector associated with the  delay pre-compensation $\kappa_{ki}$, with $\kappa_{ki}\ne \kappa_{ki'}, 1\leq i,i'\leq I_k, i \ne i'$. For ease of exposition, $\kappa_{ki}$ is an integer, and the fractional delay compensation was pursued in our previous work \cite{fDAM}. The transmit power of $\boldsymbol{\mathrm{x}}[n]$  is
\begin{align}
	&\mathbb{E}[\|\boldsymbol{\mathrm{x}}[n]\|^2]=\sum_{k=1}^K\sum_{i=1}^{{I}_k}\|\boldsymbol{\mathrm{f}}_{ki}\|^2\leq P,\label{3} \\ \notag
\end{align}
where $P$ is the maximum allowable transmit power. Let $\varphi(t)$ denote the transmit pulse shaping filter, the transmitted continuous-time signal is 
\begin{align}
	\boldsymbol{\mathrm{x}}(t)&=\sum_{n=-\infty}^{\infty}\boldsymbol{\mathrm{x}}[n] \varphi(t-nT) \label{23} \\ \notag
	&=\sum_{n=-\infty}^{\infty}\sum_{k=1}^K\sum_{i=1}^{{I}_k}\boldsymbol{\mathrm{f}}_{k{i}} s_k[n] \varphi(t-nT-\kappa_{k{i}}T) \\ \notag
	&=\sum_{k=1}^K\sum_{i=1}^{{I}_k}\boldsymbol{\mathrm{f}}_{k{i}} s_k(t-\kappa_{k{i}}T),
\end{align}
where  $\boldsymbol{\mathrm{s}}_k(t)=\sum_{n=-\infty}^{\infty} s_k[n]\varphi(t-{nT})$.

With \eqref{1} and \eqref{23}, the received signal of UE $k$  is
\begin{align}
	\boldsymbol{\mathrm{r}}_k(t)&=\boldsymbol{\mathrm{H}}_k(t)*\boldsymbol{\mathrm{x}}(t)+\boldsymbol{\mathrm{z}}_k(t) \label{1111}\\   \notag  
	&=\sum_{l=1}^{L_k}\sum_{k'=1}^{K}\sum_{i=1}^{I_{k'}} \boldsymbol{\mathrm{H}}_{kl}\boldsymbol{\mathrm{f}}_{k'i} s_{k'}(t-\kappa_{k'i}T-\tau_{kl}) + \boldsymbol{\mathrm{z}}_k(t),
\end{align}
where ${\boldsymbol{\mathrm{z}}}_k(t) \sim \mathcal{CN}({\bf{0}},{\bf{I}}_{M_r}\sigma^2)$ is  the zero-mean  additive white Gaussian noise (AWGN).

By applying the receive matched filter $\varphi(t)$ to ${\boldsymbol{\mathrm{r}}}_k(t)$, we have
\begin{align}
	&\bar{{\boldsymbol{\mathrm{r}}}}_k(t)={\boldsymbol{\mathrm{r}}}_k(t)*\varphi(-t) \label{11}\\   \notag  
	&=\sum_{l=1}^{L_k}\sum_{k'=1}^{K}\sum_{i=1}^{I_{k'}} {\boldsymbol{\mathrm{H}}}_{kl}{\boldsymbol{\mathrm{f}}}_{k'i}\sum_{n=-\infty}^{\infty}s_{k'}[n]\rho(t-{nT}-\kappa_{k'i}T-\tau_{kl})+ \bar{{\boldsymbol{\mathrm{z}}}}_k(t),
\end{align}
where $\rho(t)=\varphi(t)*\varphi(-t)$, and $\bar{{\boldsymbol{\mathrm{z}}}}_k(t)=\varphi(-t)*{\boldsymbol{\mathrm{z}}}_k(t)$ is the resulting noise after the receive matched filtering. By sampling the resulting signal $\bar{{\boldsymbol{\mathrm{r}}}}_k(t)$ at $t = n_sT$, the received signal is transformed into \eqref{1c11n}, shown at the top of next page.

\newcounter{TempEqCnt} 
\setcounter{TempEqCnt}{\value{equation}} 
\setcounter{equation}{6} 
\begin{figure*}[ht] 
	\begin{align} 
		\bar{{\boldsymbol{\mathrm{r}}}}_k[n_s]=\sum_{l=1}^{L_k}\sum_{k'=1}^{K}\sum_{i=1}^{I_{k'}} {\boldsymbol{\mathrm{H}}}_{kl}{\boldsymbol{\mathrm{f}}}_{k'i}\sum_{n=-\infty}^{\infty}s_{k'}[n]\rho((n_s-n)T-\kappa_{k'i}T-\tau_{kl}) 
		+ \bar{{\boldsymbol{\mathrm{z}}}}_k[n_s]  \label{1c11n}
	\end{align}
	\hrulefill
\end{figure*}

\newcounter{TempEqCnt1} 
\setcounter{TempEqCnt1}{\value{equation}} 
\setcounter{equation}{7} 
\begin{figure*}[ht] 
	\begin{align} 
		&y_k[n_s]=\sum_{r=1}^{R_k} {\boldsymbol{\mathrm{w}}}_{kr}^H\bar{{\boldsymbol{\mathrm{r}}}}_{k}[n_s-\mu_{kr}] =\sum_{r=1}^{R_k}\sum_{l=1}^{L_k}\sum_{k'=1}^{K}\sum_{i=1}^{I_{k'}}{\boldsymbol{\mathrm{w}}}_{kr}^H {\boldsymbol{\mathrm{H}}}_{kl}{\boldsymbol{\mathrm{f}}}_{k'i} \sum_{n=-\infty}^{\infty}s_{k'}[n] \rho((n_s-n)T-\kappa_{k'i}T-\tau_{kl}-\mu_{kr}T)  + \bar{z}_k[n_s]  \label{1c1}
	\end{align}
	\hrulefill
\end{figure*}
Unlike the existing BS-side DAM, the double-side DAM performs delay post-compensation and path-based combining at the UE, as illustrated in Fig. \ref{fig2}. Specifically, by introducing delay post-compensation $\mu_{kr}$, with $\mu_{kr}\ne \mu_{kr'}, 1\leq r,r'\leq R_k, r \ne r'$, the resulting signal after the path-based combining vector ${\boldsymbol{\mathrm{w}}}_{kr} \in \mathbb{C}^{M_r \times 1}$ is given in \eqref{1c1}, shown at the top of next page, where $\bar{z}_k[n_s]=\sum_{r=1}^{R_k} {\boldsymbol{\mathrm{w}}}_{kr}^H\bar{{\boldsymbol{\mathrm{z}}}}_k[n_s-\mu_{kr}]$.  Hence, the input-output relationship of the double-side multi-user DAM is obtained.

However, the number of delay pre-compensations, $I_k, \forall k$, and post-compensations $R_k, \forall k$, as well as the  introduced delay values $\kappa_{ki}, \mu_{kr}, \forall k,i,r$ in \eqref{1c1} need to be determined. Therefore, the subsequent discussion will focus on the design of these parameters.

\section{Double-side DAM design} \label{PD}

In this section, we mainly design $I_k, R_k, \forall k$ and  $\kappa_{ki}, \mu_{kr}, \forall k,i,r$ in \eqref{1c1}. Specifically,  we decompose this problem into two  sub-problems, namely the design for $\kappa_{ki}, \mu_{kr}, \forall k,i,r$, with fixed $I_k, R_k, \forall k$, and then design for $I_k, R_k, \forall k$.

To gain useful insights, the integer multi-path delays, i.e., $\tau_{kl}=Tn_{kl}$ are temporarily assumed, and then the received signal in \eqref{1c1} is reduced to 
\begin{align} 
 	&y_k[n_s] =\sum_{r=1}^{R_k}\sum_{l=1}^{L_k}\sum_{k'=1}^{K}\sum_{i=1}^{I_{k'}}{\boldsymbol{\mathrm{w}}}_{kr}^H {\boldsymbol{\mathrm{H}}}_{kl}{\boldsymbol{\mathrm{f}}}_{k'i} s_{k'}[n_s-\kappa_{k'i}-n_{kl}-\mu_{kr}]  + \bar{z}_k[n_s].  \label{1cn1}
\end{align}

By jointly designing the  delay pre-compensation at the BS and delay post-compensation at the UE $k$ such that  the multi-path signal components can reach UE $k$ simultaneously with delay $n_{k,\text{max}}$, the set of the desired signals of UE $k$ is then defined as
\begin{equation} 
	\mathcal{L}_k=\{i,r,l: \kappa_{ki}+\mu_{kr} + n_{kl}=n_{k,\text{max}}\} \label{bg}. 
\end{equation}
On the other hand,  the multi-path signal components not belonging to  the set $\mathcal{L}_k$ of UE $k$ are regarded as  the ISI, denoted as
\begin{equation} 
\bar{\mathcal{L}}_k=\{i,r,l: \kappa_{ki}+\mu_{kr} + n_{kl}\ne n_{k,\text{max}}\} \label{bg1}.
\end{equation} 
Hence, the received signal in \eqref{1cn1}  can be rewritten as
\begin{align}
	&y_k[n_s]=\underbrace{ \left(\sum_{l,r,i\in\mathcal{L}_k }  \boldsymbol{\mathrm{w}}_{kr}^H {\boldsymbol{\mathrm{H}}}_{kl}\boldsymbol{\mathrm{f}}_{ki}\right)s_{k}[n_s-n_{k,\text{max}}]}_{\text{Desired signal}}\label{ggw}   \\    \notag 
	&+\underbrace{\sum_{l,r,i\in\bar{\mathcal{L}}_k}{\boldsymbol{\mathrm{w}}}_{kr}^H {\boldsymbol{\mathrm{H}}}_{kl}{\boldsymbol{\mathrm{f}}}_{ki} s_{k}[n_s-\kappa_{ki}-n_{kl}-\mu_{kr}]}_{\text{ISI}}\\   \notag
	&+\underbrace{\sum_{r=1}^{R_k}\sum_{l=1}^{L_k}\sum_{k'\ne k}^{K}\sum_{i=1}^{I_{k'}}{\boldsymbol{\mathrm{w}}}_{kr}^H {\boldsymbol{\mathrm{H}}}_{kl}{\boldsymbol{\mathrm{f}}}_{k'i} s_{k'}[n_s-\kappa_{k'i}-n_{kl}-\mu_{kr}]}_{\text{IUI}} + \bar{z}_k[n_s].
\end{align}
It is observed that the first term is the desired signal, and the second and third term are ISI and IUI, respectively.  However, according to \eqref{ggw},  the sets for $\mathcal{L}_k, \forall k$  and $\bar{\mathcal{L}}_k, \forall k$ are  still unknown, and thus the design of these sets will be addressed in the following subsections.

\subsection{Constraint for the introduced delays} 
With any given $\{R_k, I_k\}_{k=1}^K$, the values of  $\kappa_{ki}, \mu_{kr}, \forall k,  i, r$  in \eqref{ggw} need to be designed. The total  possible values of UE $k$ for the introduced delays $\{\kappa_{ki}+\mu_{kr}, \forall  i, r\}$ is
\begin{align}  
	&\sum_{r=1}^{R_k}\sum_{i=1}^{I_k}(\kappa_{ki}+\mu_{kr}). \label{1bb}\\ \notag
\end{align}
For convenience, we first express the introduced delays as the product of the matrix ${\bf{Q}}_{k}\in \mathbb{C}^{I_kR_k \times (I_k+R_k)}$, and the coefficient vector $\boldsymbol{x}_{k}=[\boldsymbol{\mathrm{\kappa}}_{k}^T,\boldsymbol{\mathrm{\mu}}_k^T]^T$, with $\boldsymbol{\mathrm{\kappa}}_{k}=[\kappa_{k1},...,\kappa_{kI_k}]^T$ and $\boldsymbol{\mathrm{\mu}}_k=[\mu_{k1},...,\mu_{kR_k}]^T$, given by
\begin{equation}
	{\bf{Q}}_{k}\boldsymbol{x}_{k}= \label{DAM-1} 
	\begin{matrix} \underbrace{
			\begin{bmatrix} 
				{\bf{{B}}}_1 & {\bf{I}}_{R_k} \\
				{\bf{{B}}}_{2} & {\bf{I}}_{R_k}\\
				\vdots &\vdots \\
				{\bf{{B}}}_{I_k} & {\bf{I}}_{R_k}
		\end{bmatrix}} \\ {\bf{Q}}_{k} \end{matrix}
	\begin{matrix}  \underbrace{ 
			\begin{bmatrix} 
				\kappa_{k1}\\
				\vdots\\
				{\kappa_{kI_k}} \\
				\mu_{k1} \\
				\vdots\\
				\mu_{kR_k}
		\end{bmatrix}} \\ \boldsymbol{x}_{k} \end{matrix},
\end{equation}
where $[{\bf{{B}}}_{i}]_{:,i}\in\mathbb{C}^{R_k\times 1}={\bf{{1}}}_{R_k \times 1}, i\in [1,I_k]$, and the remaining elements of ${\bf{{B}}}_{i}$ are zero, and  $\text{rank}({\bf{Q}}_{k})\leq\text{min}\{I_kR_k,I_k+R_k\}$. To better illustrate \eqref{DAM-1},  a  example of ${\bf{Q}}_{k}$,  with $I_k=3, R_k=2, \forall k$, is expressed as
\begin{align}
	{\bf{Q}}_{k}=\begin{bmatrix} 
		1 & 0&0&\vline &1&0 \\
		1 & 0&0& \vline&0&1 \\
		\hline
		0 & 1&0& \vline&1&0 \\
		0 & 1&0&\vline& 0&1 \\
		\hline
		0 & 0&1&\vline &1&0 \\
		0 & 0&1& \vline&0&1 \\
	\end{bmatrix}=\begin{bmatrix} 
		{\bf{{B}}}_1 & {\bf{I}}_{2} \\
		{\bf{{B}}}_{2} & {\bf{I}}_{2}\\
		{\bf{{B}}}_{3} & {\bf{I}}_{2}
	\end{bmatrix},
\end{align}
and its rank is $\text{rank}({{\bf{Q}}}_{k}) = 4$.

Denote by  ${{\bf{V}}}_k\in \mathbb{C}^{{L}_k \times (I_kR_k)}$ the selection matrix. To fully exploit the ${L}_k$ multi-path signal components, ${{\bf{V}}}_k$ is designed such that 
\begin{equation} 
	{{\bf{V}}}_k{{\bf{Q}}}_{k}\boldsymbol{x}_k={\bf{n}}_k, \label{hh}
\end{equation}
where  ${\bf{n}}_k=[{n_{k,\text{max}}-n_{kL_k}},...,n_{k,\text{max}}-n_{k1}]^T$ denotes the delay alignment vector, with $n_{k,\text{max}}=n_{kL_k}$.


\begin{theorem} \label{1t1}
	For  any given ${\bf{n}}_k$, a sufficient  condition for \eqref{hh} to have a solution is 
	\begin{equation} 
		I_k+R_k-1\geq  {L}_k.\label{ppl}
	\end{equation} 
\end{theorem}

\begin{Proof}
	For any given  ${\bf{n}}_k$,  \eqref{hh} having solutions means that $\text{rank}({{\bf{V}}}_k{{\bf{Q}}}_{k})=\text{rank}([{{\bf{V}}}_k{{\bf{Q}}}_{k}, {\bf{n}}_k])= {L}_k$. Thus, the matrix ${{\bf{V}}}_k$ aims to select $ {L}_k$ linearly independent rows from ${{\bf{Q}}}_k$. 
	
	Then, we study the rank of  ${\bf{Q}}_k$. After some elementary transformations,  ${\bf{Q}}_k$ can be transformed to
	\begin{align}
		{\bf{Q}}_k \xrightarrow{}
		\begin{bmatrix} 
			{\bf{1}}_{R_k\times 1}~{\bf{0}}_{R_k\times (I_k-1)} & {\bf{I}}_{R_k} \\
			-{\bf{1}}_{(I_k-1)\times 1}~{\bf{I}}_{I_k-1} & {\bf{0}}_{(I_k-1)\times R_k} \\
			{\bf{0}}_{(I_kR_k-I_k-R_k+1)\times (I_k+R_k)}
		\end{bmatrix}  \label{DAM-11}.
	\end{align}
	It is observed that ${{\bf{Q}}}_k$ has $I_k+R_k-1$ linearly independent rows, i.e.,  $\text{rank}({{\bf{Q}}}_k)= I_k+R_k-1$.  Thus, we have	$\text{rank}({{\bf{V}}}_k{{\bf{Q}}}_{k})\leq\text{rank}({{\bf{Q}}}_k)~({L}_k \leq I_k+R_k-1)$, and then Theorem \ref{1t1} is proved.
\end{Proof}

Since the matrix ${{\bf{V}}}_k$ aims to select ${L}_k$ linearly independent rows from ${{\bf{Q}}}_k$, according to the elementary transformations in \eqref{DAM-11}, we can achieve this by selecting the first row block matrix $[{\bf{{B}}}_1  ~ {\bf{I}}_{R_k}]$ and at least one row from the remaining block matrices $[{\bf{{B}}}_{i}  ~ {\bf{I}}_{R_k}], i \in [2, I_k]$ in ${{\bf{Q}}}_k$. Consequently, ${{\bf{V}}}_k$ can be constructed as a block matrix with $L_k$ columns, where each column contains at most one ‘1’ and all other elements are zero; additionally, each row contains exactly one `1'.

In the following, we will mainly detail the design of $\kappa_{ki}, \mu_{kr},  \forall k,i,r$.

\subsection{Pre/post compensation vectors for multi-user double-side DAM} \label{lp}

For ease of exposition, let
$I_k+R_k-1=L_k$. The selection matrix ${{\bf{V}}}_k$ is designed such that the first block matrix  $[{\bf{{B}}}_1~  {\bf{I}}_{R_k}]$ and the last row of the block matrix $[{\bf{{B}}}_i~  {\bf{I}}_{R_k}], i\in [2,I_k]$, are selected, i.e., 
\begin{align}
	{{\bf{V}}}_k{{\bf{Q}}}_{k}  
	=\begin{bmatrix} 
		{\bf{{B}}}_1 & {\bf{I}}_{R_k} \\
		\bar{{\bf{{B}}}}_{I_k} & {\bf{{B}}}_{R_k}\\
	\end{bmatrix}\label{DAM-11nnl}, 
\end{align}
where $\bar{{\bf{{B}}}}_{I_k}\in\mathbb{C}^{(I_k-1)\times I_k}=[{\bf{{0}}}_{(I_k-1)\times 1}, {\bf{{I}}}_{I_k-1}]$, and ${\bf{{B}}}_{R_k}\in\mathbb{C}^{(I_k-1)\times R_k}$ follows the similar definition of ${\bf{{B}}}_{i}$ after \eqref{DAM-1}.
For example, when $L_k=4, I_{k}=3, R_k=2$, ${{\bf{V}}}_k{{\bf{Q}}}_k$ is
\begin{align}
	{{\bf{V}}}_k{{\bf{Q}}}_{k}&= \label{DAM-11nn} 
	\begin{matrix}
		\underbrace{
			\begin{bmatrix} 
				1 & 0& 0&0&0&0 \\
				0 & 1& 0&0&0&0 \\
				0 & 0& 0&1&0&0 \\
				0 & 0& 0&0&0&1 \\
		\end{bmatrix} }\\ {\bf{V}}_k
	\end{matrix}
	\begin{matrix}
		\underbrace{
			\begin{bmatrix} 
				1 & 0&0& 1&0 \\
				1 & 0&0& 0&1 \\
				0 & 1&0& 1&0 \\
				0 & 1&0& 0&1\\
				0 & 0&1& 1&0 \\
				0 & 0&1& 0&1\\
		\end{bmatrix} }\\ {\bf{Q}}_k
	\end{matrix}\\ \notag
	&=\begin{bmatrix} 
		1 & 0& 0&\vline& 1&0 \\
		1 & 0& 0&\vline& 0&1 \\
		\hline
		0 & 1& 0&\vline& 0&1\\
		0 & 0& 1&\vline& 0&1\\
	\end{bmatrix}=\begin{bmatrix} 
		{\bf{{B}}}_1 & {\bf{I}}_{2} \\
		\bar{{\bf{{B}}}}_{3} & {\bf{{B}}}_{2}\\
	\end{bmatrix}. 
\end{align}
\begin{figure}[H]
	\centering
	\includegraphics[scale=0.5]{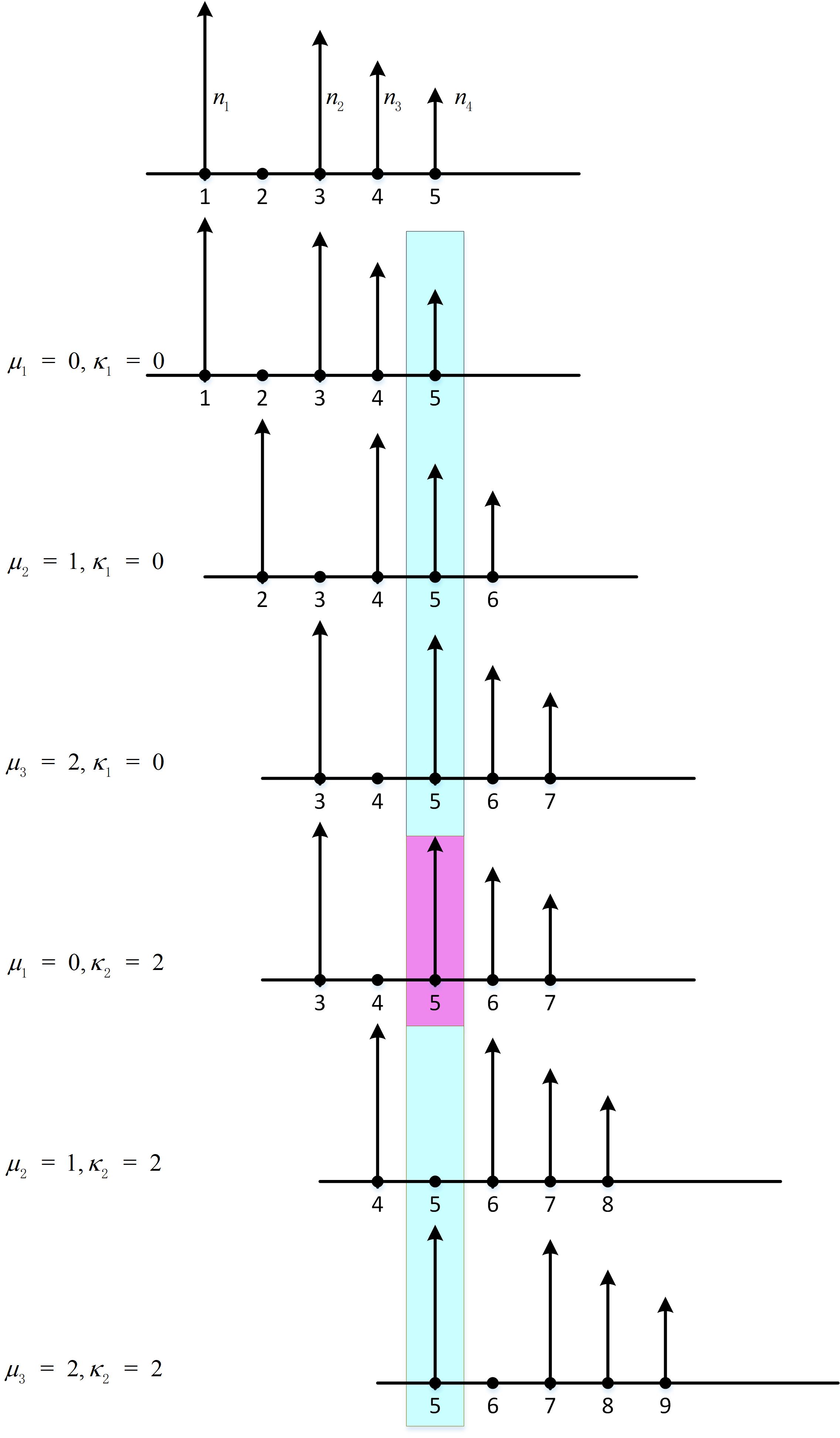}
	\caption{An example to illustrate the perfect double-side DAM with integer channel delays, where $I=2, R=3, L=4$,  $n_{1}=1, n_{2}=3, n_{3}=4$, and $n_{4}=5$.}\label{fig31}
\end{figure}

\begin{proposition} \label{1w1}
	With the given selection matrix ${{\bf{V}}}_k$ in \eqref{DAM-11nnl}, the values of the delay post-compensation and pre-compensation are 
	\begin{equation} 	
		\mu_{k(L_k+1-l)}=n_{k,\text{max}}-n_{kl}, l\in[L_k-R_k+1,L_k], \label{jj1}
	\end{equation}
	and
	\begin{equation}
		\kappa_{k(I_k+1-l)}=n_{kI_k}-n_{kl}, l\in [1,L_k-R_k+1], \label{oob}
	\end{equation} 
	respectively.
\end{proposition}

\begin{Proof}
	With $[{\bf{{B}}}_1 ~ {\bf{I}}_{R_k} ]\boldsymbol{x}_k=[{\bf{n}}_k]_{1:R_k}$, the delay post-compensation $\{\mu_{kr}\}_{r=1}^{R_k}$ is  $\kappa_{k1}+\mu_{kr}=n_{k,\text{max}}-n_{k(L_k+1-r)}, r\in[1,R_k]$.
	Without loss of generality, let $\kappa_{k1} =0$, we have
    \begin{equation}
		\mu_{kr} = n_{k,\text{max}}-n_{k(L_k+1-r)} \ge 0, r\in[1,R_k]. \label{jj}
	\end{equation}
	
	Next, the delay pre-compensation $\{\kappa_{ki}\}_{i=2}^{I_k}$ is designed based on the above obtained $\{\mu_{kr}\}_{r=1}^{R_k}$. Specifically, with   $[\bar{{\bf{{B}}}}_{I_k} ~ {\bf{{B}}}_{R_k}]\boldsymbol{x}_k=[{\bf{n}}_k]_{(R_k+1):L_k}$,  we have $\kappa_{ki}+\mu_{kR_k}=n_{k,\text{max}}-n_{k(I_k+1-i)}, i\in[2,I_k]$. 
	According to \eqref{jj},  we have  $\mu_{kR_k}=n_{k,\text{max}}-n_{kI_k}$, it then follows that  
	\begin{equation}
		\kappa_{ki}=n_{kI_k}-n_{k(I_k+1-i)}\geq 0, i\in [1, I_k] \label{D2}.
	\end{equation}
	
	Moreover, for ease of exposition, $n_{kl}, \forall k,l$ in \eqref{jj} and \eqref{D2} are re-expressed by the simple index substitution, and thus we have \eqref{jj1}  and \eqref{oob}.  As a result, it follows that 
	\begin{align}
		&n_{kl}+\kappa_{k1}+\mu_{k(L_k+1-l)}=n_{kl}+n_{k,\text{max}}-n_{kl}\label{llm} \notag \\
		&=n_{k,\text{max}}, l\in[L_k-R_k+1,L_k],
	\end{align}
	and
	\begin{align} 
		&n_{kl}+\kappa_{k(I_k+1-l)}+\mu_{kR_k}=n_{kl}+n_{k,\text{max}}-n_{kl} \label{llm1}  \notag \\
		&=n_{k,\text{max}}, l\in[1,L_k-R_k], 
	\end{align}
	which implies that there exist $L_k$ multi-path signal components with the delay $n_{k,\text{max}}$ after delay pre- and post-compensation, which are treated as the desired signals and included in $\mathcal{L}_k$.
\end{Proof}

Hence, the pre-compensation vector $\boldsymbol{\mathrm{\kappa}}_{k}$ and the post-compensation vector $\boldsymbol{\mathrm{\mu}}_{k}$ of UE $k$ are obtained, and the  sets in \eqref{bg} and \eqref{bg1} can be uniformly re-expressed by applying the following relationship
\begin{equation} 
\kappa_{ki}+\mu_{kr}=n_{k,\text{max}}-n_{k(L_k+1-r)}+n_{kI_{k'}}-n_{k(I_{k}+1-i)} \label{bgg}.
\end{equation}

It is also worth mentioning that there may exist other multi-path signal components having the resulting delay of $n_{k,\text{max}}$, with its number denoted as $L_{k,\mathrm{extra}}\geq 0$. The total number of  desired signals is $L_{k,{\mathrm{total}}}=L_{k,\mathrm{extra}}+L_k$, i.e., $|\mathcal{L}_k| = L_{k,{\mathrm{total}}}$. As an example, Fig. \ref{fig31} shows the result of the case, with $n_1=1, n_2=3, n_3=4, n_4=5$ and $n_{\text{max}} = n_{4} = 5$. For the sake of clarity, the user index $k$ is omitted. The number of delay pre-compensation is $I =2$, given by $\kappa_i=n_{2}-n_{3-i}, i \in [1,2]$, and the number of post-compensation is $R = 3$, given by $\mu_r=n_4-n_{5-r}, r\in [1,3]$. The multi-path signal components within the blue and pink boxes represent the desired signals with a delay of $n_{\text{max}}$, where the signal component in the pink box denotes an additional component with a delay of $n_{\text{max}}$.

With the delay pre- and post-compensation design given in Proposition \ref{1w1}, the resulting signal in \eqref{ggw} is rewritten as
\begin{align}
	&y_k[n_s]=\underbrace{\left(\sum_{l,r,i\in\mathcal{L}_k } {\boldsymbol{\mathrm{w}}}_{kr}^H {\boldsymbol{\mathrm{H}}}_{kl}{\boldsymbol{\mathrm{f}}}_{ki}\right)s_{k}[n_s-n_{k,\text{max}}]}_{\text{Desired signal}}\label{gg}   \\    \notag
	&+\underbrace{\sum_{l,r,i\in\bar{\mathcal{L}}_k}{\boldsymbol{\mathrm{w}}}_{kr}^H {\boldsymbol{\mathrm{H}}}_{kl}{\boldsymbol{\mathrm{f}}}_{ki} s_{k}[n_s-n_{k,\text{max}}-{\Delta}_{kkl,ri}]}_{\text{ISI}}\\   \notag
	&+\underbrace{\sum_{r=1}^{R_k}\sum_{l=1}^{L_k}\sum_{k'\ne k}^{K}\sum_{i=1}^{I_{k'}}{\boldsymbol{\mathrm{w}}}_{kr}^H {\boldsymbol{\mathrm{H}}}_{kl}{\boldsymbol{\mathrm{f}}}_{k'i} s_{k'}[n_s-n_{k,\text{max}}-{\Delta}_{kk'l,ri}]}_{\text{IUI}}  + \bar{z}_k[n_s],
\end{align}
where ${\Delta}_{kk'l,ri} = n_{kl}+n_{k'I_{k'}}-n_{k'(I_{k'}+1-i)}-n_{k(L_k+1-r)}$.

In the following, we continue to design the  $\{I_k\}_{k=1}^K$ and $\{R_k\}_{k=1}^K$ of $\kappa_{ki}, \forall k,i$ and $\mu_{kr}, \forall k,r$  for pre- and post-compensation  of double-side DAM system, respectively.

\subsection{How to choose the number of pre/post-compensation?} \label{rrrr1}

For ease of exposition, the number of  desired signals of UE $k$ in \eqref{gg} is $L_k$, and thus the maximum  number of ISI  signals for UE $k$ is $R_kL_kI_k-L_{k}$. In order to minimize the number of ISI signals from UE $k$, the problem can be formulated as 

\begin{align}  
	\min\limits_{I_k, R_k}~& R_kL_kI_k-L_{k}\label{122}\\ \notag
	\text{s.t.}& 1\leq {I}_k\leq M_t, \tag{\ref{122}{a}} \label{122a}\\ \notag
	& 1 \leq {R}_k\leq M_r,  \tag{\ref{122}{b}} \label{122b}\\ \notag
	& {I}_k+{R}_k-1={L}_k.\tag{\ref{122}{c}} \label{122c}\\ \notag
\end{align}
Since the transmitted signal and received signal comprise $I_{k}$ and $R_{k}$ path-based signals, respectively, and thus requiring that $M_t\geq I_{k}$ and $M_r\geq R_k$. Note that  with \eqref{122c}, the objective function can be expressed as  $f(I_k)=L_k(L_k+1-I_k)I_k-L_{k}$. Moreover, by applying ${I}_k+{R}_k-1={L}_k$, \eqref{122b} is re-expressed as   $L_k+1-M_r\leq {I}_k\leq  L_k$.  Thus, problem \eqref{122} is reduced to
\begin{align}  
	\min\limits_{I_k}~&f(I_k)\label{123}\\ \notag
	\text{s.t.}~& \text{max}\{1,L_k+1-M_r\}\leq {I}_k\leq \text{min}\{L_k,M_t\}. \tag{\ref{123}{a}} \label{123a}\\ \notag
\end{align}
Based on above, the optimal solution for problem \eqref{123}   can be divided into the following four cases 
\begin{itemize}
	\item Case 1: when $ M_r < L_k$ and $M_t\geq {L}_k$, we obtain $I_k= L_k$, which is referred to as BS-side DAM.
	
	\item Case 2: when $M_r\geq {L}_k$ and $M_t< {L}_k$, we obtain $I_k=1$, which  is referred to as UE-side DAM.
	
	\item Case 3: when $M_r\geq L_k$ and $M_t\geq {L}_k$, we obtain $I_k=L_k$ or $I_k=1$, which  is referred to as single-side DAM.

	\item Case 4: when $ M_r < L_k$ and $ M_t< {L}_k$,  if $f(L_k+1-M_r)\geq f(M_t)$, which is $M_r(L_k+1-M_r) \geq M_t(L_k+1-M_t)$, we obtain $I_k=M_t$; otherwise, we obtain $I_k=L_k+1-M_r$. This case  is referred to as double-side DAM.
\end{itemize}
It is observed from the above four cases that, when $M_r \geq L_k$ or $ M_t\geq {L}_k$, the single-side DAM, which includes BS-side DAM and UE-side DAM, is preferable, rather than the double-side DAM. However,  when $M_r < L_k$ and $ M_t< {L}_k$,  the double-side DAM  should be applied. Thus, depending on the physical dimensions of the transmitter and the receiver, double- and single-side DAM can be chosen, and the details are summarized in Algorithm \ref{DAM11}.

\subsection{Sum rate for eigen-beamforming transmission}
In this subsection, we propose a simple eigen-beamforming transmission scheme \cite{Eigen-1,Eigen-2} to analyze the performance of  the above cases.

Note that  the desired signal and interference signal in \eqref{gg} may have correlated terms, and the symbols with the identical delay difference need to be properly grouped. Specifically, for  ${\Delta}_{kk'l,ri}$, we have ${\Delta}_{kk'l,ri}\in \{\Delta_{kk',\text{min}},\Delta_{kk',\text{min}}+1,...,\Delta_{kk',\text{max}}\}$, with $\Delta_{kk',\text{min}}=\min\limits_{1\leq l \leq L_k}\Big(n_{kl}+n_{k'I_{k'}}\Big)-\max\limits_{1\leq i\leq I_{k'},1\leq r \leq R_k}\Big(n_{k'(I_{k'}+1-i)}+n_{k(L_k+1-r)}\Big)=n_{k1}-n_{k{L}_k}$ and $\Delta_{kk',\text{max}}=\max\limits_{1\leq l \leq L_k}\Big(n_{kl}+n_{k'I_{k'}}\Big)-\min\limits_{1\leq i\leq I_{k'},1\leq r \leq R_k}\Big(n_{k'(I_{k'}+1-i)}+n_{k(L_k+1-r)}\Big)=n_{k{L}_{k}}+n_{k'{I}_{k'}}-n_{k'1}-n_{kI_k}$. For any $q \in [\Delta_{kk',\text{min}},\Delta_{kk',\text{max}}]$, the received signal can be rewritten as

\begin{align}
	y_k[n_s]&=\sum_{r=1}^{R_k}\sum_{i=1}^{I_k}\boldsymbol{\mathrm{w}}_{kr}^H \boldsymbol{\mathrm{G}}_{kk,ri}[0]\boldsymbol{\mathrm{f}}_{ki}s_{k}[n_s-n_{k,\text{max}}]  \label{bbb} \\ \notag
	&	+\sum_{r=1}^{R_k}\sum_{i=1}^{I_k}\sum_{q=\Delta_{kk,\text{min}}, q\ne 0}^{\Delta_{kk,\text{max}}}\boldsymbol{\mathrm{w}}_{kr}^H \boldsymbol{\mathrm{G}}_{kk,ri}[q]\boldsymbol{\mathrm{f}}_{ki} s_{k}[n_s-n_{k,\text{max}}-q]\\ \notag
	&+\sum_{k'\ne k}^{K}\sum_{q=\Delta_{kk',\text{min}}}^{\Delta_{kk',\text{max}}}\sum_{r=1}^{R_k}\sum_{i=1}^{I_{k'}} \boldsymbol{\mathrm{w}}_{kr}^H \boldsymbol{\mathrm{G}}_{kk',ri}[q]\boldsymbol{\mathrm{f}}_{k'i}s_{k'}[n_s-n_{k,\text{max}}-q]+ \bar{z}_k[n_s],\\ \notag
\end{align}
where  $\boldsymbol{\mathrm{G}}_{kk',ri}[q]$ denotes the effective channel, defined as 
\begin{equation}
	\boldsymbol{\mathrm{G}}_{kk',ri}[q]  = \label{ee1v}
	\begin{cases}
		\boldsymbol{\mathrm{H}}_{kl}, &\text{if}~\exists l\in \{1,...,L_k\}, ~\mbox{s.t.}~{\Delta}_{kk'l,ri}=q,  \\ 
		\boldsymbol{\mathrm{0}} , &\mbox{otherwise}.
	\end{cases}
\end{equation}

Besides, the resulting signal to interference plus noise ratio (SINR) is given  in \eqref{ggx}, shown at the top of the next page, where $\bar{\boldsymbol{\mathrm{f}}}_{k'}=[\boldsymbol{\mathrm{f}}_{k'1}^H,...,\boldsymbol{\mathrm{f}}_{k'I_{k'}}^H]^H \in \mathbb{C}^{M_tI_{k'} \times 1}$, $\bar{\boldsymbol{\mathrm{w}}}_{k}=[\boldsymbol{\mathrm{w}}_{k1}^H,...,\boldsymbol{\mathrm{w}}_{kR_{k}}^H]^H \in \mathbb{C}^{M_rR_{k} \times 1}$, and $\bar{\boldsymbol{\mathrm{G}}}_{kk'}[q]\in \mathbb{C}^{M_rR_k \times M_t I_{k'}}$ is
\newcounter{TempEqCnt7} 
\setcounter{TempEqCnt7}{\value{equation}} 
\setcounter{equation}{32} 
\begin{figure*}[ht] 
	\begin{align}
		\gamma_k=\frac{|\bar{\boldsymbol{\mathrm{w}}}_{k}^H \bar{\boldsymbol{\mathrm{G}}}_{kk}[0]\bar{\boldsymbol{\mathrm{f}}}_{k}|^2}{\sum_{q=\Delta_{kk,\text{min}}, q\ne 0}^{\Delta_{kk,\text{max}}}|\bar{\boldsymbol{\mathrm{w}}}_{k}^H \bar{\boldsymbol{\mathrm{G}}}_{kk}[q]\bar{\boldsymbol{\mathrm{f}}}_{k}|^2+\sum_{k'\ne k}^{K}\sum_{q=\Delta_{kk',\text{min}}}^{\Delta_{kk',\text{max}}}|\bar{\boldsymbol{\mathrm{w}}}_{k}^H \bar{\boldsymbol{\mathrm{G}}}_{kk'}[q]\bar{\boldsymbol{\mathrm{f}}}_{k'}|^2+\sigma^2\|\bar{\boldsymbol{\mathrm{w}}}_{k}\|^2}. \label{ggx}
	\end{align}
	\hrulefill
\end{figure*}

\begin{align}
	&\bar{\boldsymbol{\mathrm{G}}}_{kk'}[q]= \begin{bmatrix}\boldsymbol{\mathrm{G}}_{kk',11}[q] &\boldsymbol{\mathrm{G}}_{kk',12}[q]& \cdots &\boldsymbol{\mathrm{G}}_{kk',1I_{k'}}[q]\label{s-2}\\
		\boldsymbol{\mathrm{G}}_{kk',21}[q]  &\boldsymbol{\mathrm{G}}_{kk',22}[q]& \cdots &\boldsymbol{\mathrm{G}}_{kk',2I_{k'}}[q]\\
		\vdots &\vdots & \ddots &\vdots\\
		\boldsymbol{\mathrm{G}}_{kk',R_k1}[q]& \boldsymbol{\mathrm{G}}_{kk',R_k2}[q]& \cdots &\boldsymbol{\mathrm{G}}_{kk',R_kI_{k'}}[q] \end{bmatrix}. \\ \notag
\end{align}

Let the (reduced) singular value decomposition (SVD) of $\bar{\boldsymbol{\mathrm{G}}}_{kk}[0]$ be $\bar{\boldsymbol{\mathrm{G}}}_{kk}[0]= {{\bf{U}}}_{k}{{\bf{\Sigma}}}_{k}{{\bf{V}}}_{k}^H$, where ${{\bf{U}}}_{k}\in \mathbb{C}^{M_rR_k \times r}$,  ${{\bf{\Sigma}}}_{k}\in \mathbb{C}^{r \times r}$, with the singular values in a descending order, and ${{\bf{V}}}_{k}\in \mathbb{C}^{M_t I_k \times r}$. Besides, let  ${{\bf{u}}}_{k}=[{{\bf{U}}}_{k}]_{:,1}$ and ${{\bf{v}}}_{k}=[{{\bf{V}}}_{k}]_{:,1}$  denote the left and right singular vectors corresponding to the largest singular value. Thus, the transmit beamforming  and receive combining vectors for the eigen-beamforming transmission are $\bar{{\boldsymbol{\mathrm{f}}}}_{k}=\sqrt{P}{{\bf{v}}}_{k}/\|{\bf{V}}_{f}\|_{\text{F}}$ with ${\bf{V}}_{f}=[{\bf{v}}_{1},...,{\bf{v}}_{K}]$, and $\bar{{\boldsymbol{\mathrm{w}}}}_{k}={{\bf{u}}}_{k}/\|{\bf{u}}_{k}\|$, respectively. By substituting $\bar{{\boldsymbol{\mathrm{f}}}}_{k}$ and $\bar{{\boldsymbol{\mathrm{w}}}}_{k}$ into \eqref{ggx}, the achievable rate of UE $k$ is obtained, given by $R_k=\text{log}_2(1+\gamma_k)$. Thus, the sum rate is $\sum_{k=1}^KR_k$.

It is also worth mentioning that the double-side DAM includes the single-side DAM as a special case, rendering the aforementioned transmission scheme applicable to the single-side DAM. Moreover, considering the developing trends of  XL-MIMO and mmWave/THz communications in future 6G networks, the four cases demonstrate that when the number of BS/UE antennas is sufficiently large, single-side DAM, where delay compensation is only performed at the BS/UE, is preferred than double-side DAM. This is expected since less  ISI needs to be spatially eliminated in single-side DAM. Therefore, the single-side DAM is considered for further analysis. Moreover, BS-side DAM is more suitable for downlink systems. Thus, the subsequent analysis focuses on BS-side DAM, taking into account the general fractional channel delays.

\section{Sum rate for BS-side multi-user DAM and benchmarking scheme  with fractional channel delays} \label{gggg}
In this section, we examine the sum rate of the multi-user BS-side DAM system by evaluating two low-complexity beamforming schemes: eigen-beamforming and ISI-ZF. For performance comparison, we also consider the benchmarking scheme of OFDM.

\subsection{BS-side multi-user DAM}
\subsubsection{Path-based eigen-beamforming transmission }\label{ppp}
With $I_k={L}_k$ and $R_k=1$, by letting $\kappa_{ki}=n_{k,\text{max}}-n_{ki}, \forall k$ and sampling at $t = n_sT + n_{k,\text{max}}T$  for $\bar{\boldsymbol{\mathrm{r}}}_k(t)$ in \eqref{11}, the received signal in \eqref{1c1} reduces to
	\begin{align} 
	&y_k[n_s] =\underbrace{\sum_{l=1}^{L_k}{\boldsymbol{\mathrm{w}}}_{k}^H {\boldsymbol{\mathrm{H}}}_{kl}{\boldsymbol{\mathrm{f}}}_{kl} \rho_{kk,ll}[0]s_{k}[n_s]}_{\text{Desired signal}}+\underbrace{\sum_{l=1}^{L_k}{\boldsymbol{\mathrm{w}}}_{k}^H {\boldsymbol{\mathrm{H}}}_{kl}{\boldsymbol{\mathrm{f}}}_{kl}  \sum_{n\ne n_s}\rho_{kk,ll}[n_s-n]s_{k}[n]}_{\text{ISI}} \label{nnnv}\notag \\  
	& +\underbrace{\sum_{l=1}^{L_k}\sum_{i\ne l}^{L_{k}}{\boldsymbol{\mathrm{w}}}_{k}^H {\boldsymbol{\mathrm{H}}}_{kl}{\boldsymbol{\mathrm{f}}}_{ki}  \sum_{n=-\infty}^{\infty}\rho_{kk,li}[n_s-n]s_{k}[n]}_{\text{ISI}}\\ \notag
	& +\underbrace{\sum_{l=1}^{L_k}\sum_{k'\ne k}^{K}\sum_{i=1}^{L_{k'}}{\boldsymbol{\mathrm{w}}}_{k}^H {\boldsymbol{\mathrm{H}}}_{kl}{\boldsymbol{\mathrm{f}}}_{k'i}  \sum_{n=-\infty}^{\infty}\rho_{kk',li}[n_s-n]s_{k'}[n]}_{\text{IUI}} + \bar{z}_k[n_s],
\end{align}
where $\rho_{kk',li}[n_s-n]=\rho\big((n_s-n)T + n_{k,\text{max}}T-n_{k',\text{max}}T-{\tau}_{\text{F},kl}-n_{kl}T+n_{k'i}T\big)$.  Hence, the  power  in the first and second (FS) parts of  \eqref{nnnv} can be expressed as
\begin{equation}
	\begin{split}
		P_{\text{FS}}(\boldsymbol{\mathrm{w}}_k,\bar{\boldsymbol{\mathrm{f}}}_{k})&=\sum_{n=-\infty}^{\infty}\left|\sum_{l=1}^{L_k}{\boldsymbol{\mathrm{w}}}_{k}^H {\boldsymbol{\mathrm{H}}}_{kl}\rho_{kk,ll}[n_s-n]{\boldsymbol{\mathrm{f}}}_{kl} \right|^2\\ 
		&=\sum_{n=-\infty}^{\infty}|{\boldsymbol{\mathrm{w}}}_{k}^H{{\boldsymbol{\mathrm{H}}}}_{k,\rho}[n_s-n]\bar{{\boldsymbol{\mathrm{f}}}}_{k}|^2,\label{DS1}
	\end{split}
\end{equation}
where ${{\boldsymbol{\mathrm{H}}}}_{k,\rho}[n_s-n]=\left[{\boldsymbol{\mathrm{H}}}_{k1}\rho_{kk,11}[n_s-n],...,{\boldsymbol{\mathrm{H}}}_{kL_k}\rho_{kk,L_kL_k}[n_s-n]\right]\in \mathbb{C}^{M_r \times M_t L_k}$. Then,  the desired signal (DS) power  of \eqref{DS1} is
\begin{equation}
	\begin{split}
		P_{\text{DS}}(\boldsymbol{\mathrm{w}}_k,\bar{\boldsymbol{\mathrm{f}}}_{k})&=|{\boldsymbol{\mathrm{w}}}_{k}^H {\boldsymbol{\mathrm{H}}}_{k,\rho}[0]\bar{\boldsymbol{\mathrm{f}}}_{k}|^2, \label{DS}
	\end{split}
\end{equation}
and the ISI part is
\begin{equation}
	\begin{split}
		P_{\text{ISI,1}}(\boldsymbol{\mathrm{w}}_k,\bar{\boldsymbol{\mathrm{f}}}_{k})&=\sum_{n\ne n_s}|{\boldsymbol{\mathrm{w}}}_{k}^H {\boldsymbol{\mathrm{H}}}_{k,\rho}[n_s-n]\bar{\boldsymbol{\mathrm{f}}}_{k}|^2.
	\end{split}
\end{equation}
 Besides, the power of the third part in \eqref{nnnv}  is
\begin{equation}
	\begin{split}
		P_{\text{ISI,2}}(\boldsymbol{\mathrm{w}}_k,\bar{\boldsymbol{\mathrm{f}}}_{k})&=\sum_{n=-\infty}^{\infty}\left|\sum_{l=1}^{L_k}\sum_{i\ne l}^{L_{k}}{\boldsymbol{\mathrm{w}}}_{k}^H {\boldsymbol{\mathrm{H}}}_{kl}\rho_{kk,li}[n_s-n]{\boldsymbol{\mathrm{f}}}_{ki}  \right|^2 \label{lll} \\
		&=\sum_{n=-\infty}^{\infty}\left|\sum_{i=1}^{L_k}{\boldsymbol{\mathrm{w}}}_{k}^H \hat{\boldsymbol{\mathrm{H}}}_{ki,\rho}[n_s-n]{\boldsymbol{\mathrm{f}}}_{ki}  \right|^2\\
		&=\sum_{n=-\infty}^{\infty}|{\boldsymbol{\mathrm{w}}}_{k}^H\hat{\boldsymbol{\mathrm{H}}}_{k,\rho}[n_s-n]\bar{\boldsymbol{\mathrm{f}}}_{k}|^2,
	\end{split}
\end{equation}
where $\hat{\boldsymbol{\mathrm{H}}}_{ki,\rho}[n_s-n]\in \mathbb{C}^{M_r \times M_t }=\bar{\boldsymbol{\mathrm{H}}}_{ki,\rho}[n_s-n]\bar{{\boldsymbol{\mathrm{I}}}}_{M_t} $, with  $\bar{\boldsymbol{\mathrm{H}}}_{ki,\rho}[n_s-n]=\big[{\boldsymbol{\mathrm{H}}}_{k1}\rho_{kk,1i}[n_s-n],...,{\boldsymbol{\mathrm{H}}}_{k(i-1)}\rho_{kk,(i-1)i}[n_s-n],{\boldsymbol{\mathrm{H}}}_{k(i+1)}\rho_{kk,(i+1)i}[n_s-n],...,{\boldsymbol{\mathrm{H}}}_{kL_k}\rho_{kk,L_ki}[n_s-n]\big]\in \mathbb{C}^{M_r \times M_t(L_k-1) }$, and $\bar{{\boldsymbol{\mathrm{I}}}}_{M_t}=[{{\boldsymbol{\mathrm{I}}}}_{M_t},...,{{\boldsymbol{\mathrm{I}}}}_{M_t}]^H\in \mathbb{C}^{M_t (L_k-1) \times M_t }$, and $\hat{\boldsymbol{\mathrm{H}}}_{k,\rho}[n_s-n]=\big[\hat{\boldsymbol{\mathrm{H}}}_{k1,\rho}[n_s-n],...,\hat{\boldsymbol{\mathrm{H}}}_{kL_k,\rho}[n_s-n]\big]\in \mathbb{C}^{M_r \times M_tL_k}$. Moreover,  the power of the IUI for the forth part in \eqref{nnnv} is
\begin{equation}
	\begin{split}
		P_{\text{IUI}}(\boldsymbol{\mathrm{w}}_k,\sum_{k'\ne k}\bar{\boldsymbol{\mathrm{f}}}_{k'})&=\sum_{n=-\infty}^{\infty}\sum_{k'\ne k}^{K}\left|\sum_{l=1}^{L_k}\sum_{i=1}^{L_{k'}}{\boldsymbol{\mathrm{w}}}_{k}^H {\boldsymbol{\mathrm{H}}}_{kl}\rho_{kk',li}[n_s-n]{\boldsymbol{\mathrm{f}}}_{k'i}   \right|^2 \\
		&=\sum_{k'\ne k}^{K}\sum_{n=-\infty}^{\infty}\left|{\boldsymbol{\mathrm{w}}}_{k}^H \hat{\boldsymbol{\mathrm{H}}}_{kk',\rho}[n_s-n]\bar{\boldsymbol{\mathrm{f}}}_{k'}\right|^2, \label{IUI}
	\end{split}
\end{equation}
where $\hat{\boldsymbol{\mathrm{H}}}_{kk',\rho}[n_s-n]=\bar{\boldsymbol{\mathrm{H}}}_{k}\boldsymbol{\mathrm{\rho}}_{kk'}[n_s-n]$, 
  $\bar{\boldsymbol{\mathrm{H}}}_{k}=[{\boldsymbol{\mathrm{H}}}_{k1},...,{\boldsymbol{\mathrm{H}}}_{kL_k}]$, and $\boldsymbol{\mathrm{\rho}}_{kk'}[n_s-n]=\bar{\boldsymbol{\mathrm{\rho}}}_{kk'}[n_s-n] \otimes {{\boldsymbol{\mathrm{I}}}}_{M_t} \in \mathbb{C}^{M_tL_k \times M_tL_{k'} }$, with 
\begin{align}
  	&\bar{\boldsymbol{\mathrm{\rho}}}_{kk'}[n_s-n]= \begin{bmatrix}\rho_{kk',11}[n_s-n] & \cdots &\rho_{kk',1L_{k'}}[n_s-n]\label{s-2}\\
  		\vdots  & \ddots &\vdots\\
  		\rho_{kk',L_k1}[n_s-n]&\cdots &\rho_{kk',L_kL_{k'}}[n_s-n] \end{bmatrix}. \\ \notag
  \end{align}
With \eqref{DS}-\eqref{IUI}, the resulting SINR of \eqref{nnnv} is  
\begin{align}
 	\gamma_k^{\text{BS}}&=\frac{P_{\text{DS}}(\boldsymbol{\mathrm{w}}_k, \bar{\boldsymbol{\mathrm{f}}}_{k})}{P_{\text{ISI,1}}(\boldsymbol{\mathrm{w}}_k,\bar{\boldsymbol{\mathrm{f}}}_{k})+P_{\text{ISI,2}}(\boldsymbol{\mathrm{w}}_k,\bar{\boldsymbol{\mathrm{f}}}_{k})+P_{\text{IUI}}(\boldsymbol{\mathrm{w}}_k,\sum_{k'\ne k}\bar{\boldsymbol{\mathrm{f}}}_{k'})+\sigma^2\|\boldsymbol{\mathrm{w}}_k\|^2}.\label{005}\\ \notag  
\end{align}
Similar to the eigen-beamforming for double-side DAM, with ${\boldsymbol{\mathrm{H}}}_{k,\rho}[0]= \bar{{\bf{U}}}_{k}\bar{{\bf{\Sigma}}}_{k}\bar{{\bf{V}}}_{k}^H$, let 
$\bar{{\bf{u}}}_{k}=[\bar{{\bf{U}}}_{k}]_{:,1}$ and $\bar{{\bf{v}}}_{k}=[\bar{{\bf{V}}}_{k}]_{:,1}$  denote the left and right singular vectors corresponding to the largest singular value. Thus, the transmit beamforming and receive combining vectors for the eigen-beamforming transmission are $\bar{{\boldsymbol{\mathrm{f}}}}_{k}=\sqrt{P}\bar{{\bf{v}}}_{k}/\|\bar{\bf{V}}_{f}\|_{\text{F}}$ with $\bar{\bf{V}}_{f}=[\bar{{\bf{v}}}_{1},...,\bar{\bf{v}}_{K}]$, and ${{\boldsymbol{\mathrm{w}}}}_{k}=\bar{{\bf{u}}}_{k}/\|\bar{\bf{u}}_{k}\|$, respectively, and then the resulting sum rate  can be  obtained.

\begin{algorithm}[t]
	\renewcommand{\algorithmicrequire}{\textbf{Input:}}
	\renewcommand{\algorithmicensure}{\textbf{Output:}}
	\caption{Double-side DAM design}
	\label{DAM11}
	\begin{algorithmic}[1]
		\Require $L_k, \forall k$, $n_{kl}, \forall k,l$, ${\bf{n}}_k$, $M_t$, and $M_r$. 
		\State Obtain the  constraint for the introduced delays $I_k+R_k-1 \ge L_k$ based on Theorem \ref{1t1}.
		\State Obtain the delay  post-compensation $\mu_{kr}, \forall k, r$ and pre-compensation $\kappa_{ki}, \forall k,i$, with $I_k+R_k-1 = L_k$  for simplicity, based on Proposition \ref{1w1}.
		\If {$M_r < L_k$ and $M_t \ge L_k$}
		\State $I_k= L_k;$
		\ElsIf {{$M_r\geq {L}_k$ and $M_t< {L}_k$}}
		\State $I_k=1;$
		\ElsIf {{$M_r\geq L_k$ and $M_t\geq {L}_k$}}
		\State $I_k=L_k$ or $I_k=1$;
		\ElsIf {$ M_r < L_k$ and $ M_t< {L}_k$}
		\If  {$M_r(L_k+1-M_r) \geq M_t(L_k+1-M_t)$}
		\State $I_k=M_t$;
		\Else
		\State $I_k=L_k+1-M_r$;
		\EndIf
		\EndIf
		\Ensure $\kappa_{ki}, \mu_{kr}, \forall k,i, r$, and $I_k, R_k, \forall k$. 
	\end{algorithmic}
\end{algorithm}
\subsubsection{Path-Based ISI-ZF Beamforming}\label{DAM1}
From  \eqref{nnnv}, it is observed that if \({\boldsymbol{\mathrm{w}}}_{k}, \forall k\) and \({\boldsymbol{\mathrm{f}}}_{ki}, \forall k,l\) are designed to eliminate the partial ISI and IUI, given by
\begin{align}
	&{\boldsymbol{\mathrm{w}}}_{k}^H{\boldsymbol{\mathrm{H}}}_{kl}{\boldsymbol{\mathrm{f}}}_{ki}=0,  \forall k, i\ne l,\label{ZF1vv}\notag \\ 
	&{\boldsymbol{\mathrm{w}}}_{k}^H {\boldsymbol{\mathrm{H}}}_{kl}{\boldsymbol{\mathrm{f}}}_{k'i}=0, \forall  k' \ne k, \\  \notag 
\end{align}
the received signal in \eqref{nnnv} is simplified as
\begin{align} 
	y_k[n_s] &=\sum_{l=1}^{L_k}{\boldsymbol{\mathrm{w}}}_{k}^H {\boldsymbol{\mathrm{H}}}_{kl}{\boldsymbol{\mathrm{f}}}_{kl} \rho_{kk,ll}[0]s_{k}[n_s]\label{88vv}\\ \notag 
	&+\sum_{l=1}^{L_k}{\boldsymbol{\mathrm{w}}}_{k}^H {\boldsymbol{\mathrm{H}}}_{kl}{\boldsymbol{\mathrm{f}}}_{kl}  \sum_{n\ne n_s}\rho_{kk,ll}[n_s-n]s_{k}[n]+\bar{z}_k[n_s] .
\end{align}
It is observed from \eqref{88vv} that  the residual ISI still exists for the desired symbol sequence \(s_{k}[n_s], \forall k\) due to the fractional channel delays \({\tau}_{\text{F},kl}, \forall k,l\).

\begin{proposition} \label{1vv1}
	A necessary condition for the ZF conditions in \eqref{ZF1vv} to be feasible is
	\begin{align}
		&L_kM_t+M_r\geq L_kL_{\mathrm{tot}}+1\label{eeee} \\ \notag 
	\end{align}
\end{proposition}

\begin{Proof}
	
	Let $N_e$ and $N_v$ denote the the total number of equations and the free design variables, respectively. To determine the number of equations for feasibility, we have
	\begin{align}
		&N_e=L_kL_{\mathrm{tot}}-L_k.
	\end{align}
	where $L_{\mathrm{tot}}=\sum_{k=1}^KL_k$.
	By applying a similar approach as in \cite{IA-2,IA-1,DDAM} to count the free design variables, the total number of independent variables in the system is
	\begin{equation}
		N_v=L_k(M_t-1)+(M_r-1).
	\end{equation}
	Thus, the necessary condition requires \( N_v \geq N_e \). Therefore, Proposition \ref{1vv1} holds.
\end{Proof}

Note that for the  multiple-input single-output (MISO) communication system, i.e., \(M_r=1\), \eqref{eeee} reduces to \(M_t \geq L_{\mathrm{tot}}\) as revealed in \cite{Multi-UserDAM}. 

\begin{proposition} \label{1v3}
	A sufficient conditions for the ZF constraints in \eqref{ZF1vv} to be feasible is
	\begin{align}
		&M_t\geq M_r(L_{\mathrm{tot}}-1)+1. \label{eee2} \\ \notag 
	\end{align}
\end{proposition}

\begin{Proof}
	The constraint in \eqref{ZF1vv} can be compactly written as
	\begin{align}
		& \bar{{\boldsymbol{\mathrm{H}}}}_{kl}{\boldsymbol{\mathrm{f}}}_{kl} ={\bf{0}}_{ M_r(L_{\mathrm{tot}}-1)\times 1}, \forall l, \label{ZFg1v}
	\end{align}
	where $\bar{{\boldsymbol{\mathrm{H}}}}_{kl}=[{\boldsymbol{\mathrm{H}}}_{11}^H,...,{\boldsymbol{\mathrm{H}}}_{1L_1}^H,...,{\boldsymbol{\mathrm{H}}}_{k1}^H,...,{\boldsymbol{\mathrm{H}}}_{k(l-1)}^H,{\boldsymbol{\mathrm{H}}}_{k(l+1)}^H,...,{\boldsymbol{\mathrm{H}}}_{K1}^H,...,{\boldsymbol{\mathrm{H}}}_{KL_K}^H]^H\\
	 \in \mathbb{C}^{M_r(L_{\mathrm{tot}}-1) \times M_t  }$. The above ZF constraint is feasible almost surely when $M_t\geq M_r(L_{\mathrm{tot}}-1)+1$.
\end{Proof}

According to \eqref{ZFg1v}, the  path-based transmit beamforming is designed as ${\boldsymbol{\mathrm{f}}}_{kl}=\bar{{\boldsymbol{\mathrm{H}}}}_{kl}^{\perp}{\bf{b}}_{kl}, \forall l$, where $\bar{{\boldsymbol{\mathrm{H}}}}_{kl}^{\perp}\in \mathbb{C}^{M_t\times N_t}$ is an orthogonal basis for the orthogonal complement of $\bar{{\boldsymbol{\mathrm{H}}}}_{kl}$, with $N_t=\text{rank}(\bar{{\boldsymbol{\mathrm{H}}}}_{kl}^{\perp})$, and ${\bf{b}}_{kl}\in \mathbb{C}^{N_t\times 1}$  is the transmit beamforming vector to be designed. By substituting ${\boldsymbol{\mathrm{f}}}_{kl}=\bar{{\boldsymbol{\mathrm{H}}}}_{kl}^{\perp}{\bf{b}}_{kl}$ into \eqref{88vv}, we have
\begin{align}
	&y_k[n_s] =\sum_{l=1}^{L_k}{\boldsymbol{\mathrm{w}}}_{k}^H {\boldsymbol{\mathrm{H}}}_{kl}\bar{{\boldsymbol{\mathrm{H}}}}_{kl}^{\perp}{\bf{b}}_{kl} \rho_{kk,ll}[0]s_{k}[n_s]\label{881} \\ \notag 
	&+\sum_{l=1}^{L_k}{\boldsymbol{\mathrm{w}}}_{k}^H {\boldsymbol{\mathrm{H}}}_{kl}\bar{{\boldsymbol{\mathrm{H}}}}_{kl}^{\perp}{\bf{b}}_{kl}  \sum_{n\ne n_s}\rho_{kk,ll}[n_s-n]s_{k}[n]+\bar{z}_k[n_s]\\ \notag 
	&=
	{\boldsymbol{\mathrm{w}}}_{k}^H\tilde{{\boldsymbol{\mathrm{H}}}}_{k, \rho}[0]\bar{{\bf{b}}}_k s_{k}[n_s] +\sum_{n\ne n_s}{\boldsymbol{\mathrm{w}}}_{k}^H \tilde{{\boldsymbol{\mathrm{H}}}}_{k, \rho}[n_s-n]\bar{{\bf{b}}}_ks_{k}[n] + \bar{z}_k[n_s], \\ \notag
\end{align}
where $\tilde{{\boldsymbol{\mathrm{H}}}}_{k, \rho}[n_s-n]=\left[{\boldsymbol{\mathrm{H}}}_{k1}\bar{{\boldsymbol{\mathrm{H}}}}_{k1}^{\perp}\rho_{kk,11}[n_s-n],...,{\boldsymbol{\mathrm{H}}}_{kL_k}\bar{{\boldsymbol{\mathrm{H}}}}_{kL_k}^{\perp}\rho_{kk,L_kL_k}[n_s-n]\right]\in \mathbb{C}^{M_r\times N_rL_k } $, and $\bar{{\bf{b}}}_k=[{\bf{b}}_{k1}^H,...,{\bf{b}}_{kL_k}^H]^H\in \mathbb{C}^{N_rL_k\times 1}$. Thus,  the resulting SINR of UE $k$ is
\begin{align}
	\gamma_k^{\text{ISI-ZF}}=\frac{|{\boldsymbol{\mathrm{w}}}_k^H \tilde{{\boldsymbol{\mathrm{H}}}}_{k, \rho}[0]\bar{{\bf{b}}}_k|^2}{\sum_{n\ne n_s}|{\boldsymbol{\mathrm{w}}}_{k}^H\tilde{{\boldsymbol{\mathrm{H}}}}_{k, \rho}[n_s-n]\bar{{\bf{b}}}_k|^2+\sigma^2\|{\boldsymbol{\mathrm{w}}}_k\|^2 }.
\end{align}

We aim to maximize the sum rate by  jointly optimizing the transmit beamforming vector \( \bar{{\bf{b}}}_k, \forall k \) and the receive combining vector \( {{\boldsymbol{\mathrm{w}}}}_k, \forall k\). Hence, we have
the following problem
\begin{align}
	\max\limits_{\bar{{\bf{b}}}_k, {{\boldsymbol{\mathrm{w}}}}_k, \forall k} ~&\sum_{k=1}^K\text{log}_2\left(1+\frac{|{\boldsymbol{\mathrm{w}}}_k^H \tilde{{\boldsymbol{\mathrm{H}}}}_{k, \rho}[0]\bar{{\bf{b}}}_k|^2}{\sum_{n\ne n_s}|{\boldsymbol{\mathrm{w}}}_{k}^H\tilde{{\boldsymbol{\mathrm{H}}}}_{k, \rho}[n_s-n]\bar{{\bf{b}}}_k|^2+\sigma^2\|{\boldsymbol{\mathrm{w}}}_k\|^2 }\right)\label{lblvvv} \\  \notag
	\text{s.t.}& \sum_{k=1}^K\|\bar{{\bf{b}}}_k\|^2\leq P. \tag{\ref{lblvvv}{a}} \label{lblva}\\ \notag
\end{align}
where $\|{\boldsymbol{\mathrm{f}}}_{kl}\|^2=\|\bar{{\boldsymbol{\mathrm{H}}}}_{kl}^{\perp}{\bf{b}}_{kl}\|=\|{{\bf{b}}}_{kl}\|^2, \forall k,l$.

For the given  transmit beamforming $\bar{{\bf{b}}}_k, \forall k$, problem \eqref{lblvvv} is transformed into
\begin{align}
	\max\limits_{ {{\boldsymbol{\mathrm{w}}}}_k, \forall k} ~&\sum_{k=1}^K\text{log}_2\left(1+\frac{{\boldsymbol{\mathrm{w}}}_{k}^H\tilde{{\boldsymbol{\mathrm{H}}}}_{k,\rho}[0]\bar{{\bf{b}}}_k\bar{{\bf{b}}}_k^H\tilde{{\boldsymbol{\mathrm{H}}}}_{k,\rho}[0]{{\boldsymbol{\mathrm{w}}}}_{k}}{\boldsymbol{\mathrm{w}}_{k}^H{\boldsymbol{\mathrm{\Lambda}}}_{k}^{\text{ZF}}\boldsymbol{\mathrm{w}}_{k}}\right),\label{lblvvv2} \\  \notag
\end{align}
where ${\boldsymbol{\mathrm{\Lambda}}}_{k}^{\text{ZF}}=\sum_{n\ne n_s}\tilde{{\boldsymbol{\mathrm{H}}}}_{k, \rho}[n_s-n]\bar{{\bf{b}}}_k\bar{{\bf{b}}}_k^H\tilde{{\boldsymbol{\mathrm{H}}}}_{k, \rho}^H[n_s-n]+\sigma^2{{\bf{I}}}_{M_r}$. The above problem  can be maximized by the minimum mean squared error (MMSE) beamforming \cite{MSE-R2}, given by
\begin{align}
	{{\boldsymbol{\mathrm{w}}}}_k=\frac{({\boldsymbol{\mathrm{\Lambda}}}_{k}^{\text{ZF}})^{-1}\tilde{{\boldsymbol{\mathrm{H}}}}_{k,\rho}[0]\bar{{\bf{b}}}_k}{\|({\boldsymbol{\mathrm{\Lambda}}}_{k}^{\text{ZF}})^{-1}\tilde{{\boldsymbol{\mathrm{H}}}}_{k,\rho}[0]\bar{{\bf{b}}}_k\|}. \label{rd11}
\end{align}

On the other hand, for the given receive combining vector ${{\boldsymbol{\mathrm{w}}}}_k, \forall k$, the sub-problem of optimizing the transmit beamforming vector is given by
\begin{align}
	\max\limits_{\bar{{\bf{b}}}_k, \forall k} ~&\sum_{k=1}^K\text{log}_2\left(1+\frac{|{\boldsymbol{\mathrm{w}}}_k^H \tilde{{\boldsymbol{\mathrm{H}}}}_{k, \rho}[0]\bar{{\bf{b}}}_k|^2}{\sum_{n\ne n_s}|{\boldsymbol{\mathrm{w}}}_{k}^H\tilde{{\boldsymbol{\mathrm{H}}}}_{k, \rho}[n_s-n]\bar{{\bf{b}}}_k|^2+\sigma^2\|{\boldsymbol{\mathrm{w}}}_k\|^2 }\right) \label{lblvvv1} \\  \notag
	\text{s.t.}& \eqref{lblva}.\\ \notag
\end{align} 
Note that the problem  \eqref{lblvvv1} can be solved using successive convex approximation (SCA) \cite{MU-DAM}. However, this approach will lead to high computational complexity. As an alternative, we consider the low-complexity equal power allocation scheme, with \( P_k = P/K \). Consequently, the problem  \eqref{lblvvv1} is reduced to

\begin{align}
	\max\limits_{\bar{{\bf{b}}}_k, \forall k} ~&\sum_{k=1}^K\text{log}_2\bigg(1+
	\frac{\bar{{\bf{b}}}_k^H\tilde{{\boldsymbol{\mathrm{H}}}}_{k, \rho}^H[0]{\boldsymbol{\mathrm{w}}}_k{\boldsymbol{\mathrm{w}}}_k^H \tilde{{\boldsymbol{\mathrm{H}}}}_{k, \rho}[0]\bar{{\bf{b}}}_k}{\bar{{\bf{b}}}_k^H\bar{\boldsymbol{\mathrm{\Lambda}}}_{k}^{\text{ZF}}\bar{{\bf{b}}}_k }\bigg) \label{lblvvv11} \\  \notag
		\text{s.t.}& \|\bar{{\bf{b}}}_k\|^2\leq P_k, \forall k, \tag{\ref{lblvvv11}{a}} \label{lblvaa}\\ \notag
\end{align}
where $\bar{\boldsymbol{\mathrm{\Lambda}}}_{k}^{\text{ZF}}=\sum_{n\ne n_s}\tilde{{\boldsymbol{\mathrm{H}}}}_{k, \rho}^H[n_s-n]{\boldsymbol{\mathrm{w}}}_{k}{\boldsymbol{\mathrm{w}}}_{k}^H\tilde{{\boldsymbol{\mathrm{H}}}}_{k, \rho}[n_s-n]+\sigma^2\frac{K}{P}\|{\boldsymbol{\mathrm{w}}}_k\|^2$. Thus, the MMSE transmit beamforming  is
\begin{align}
	{{\boldsymbol{\mathrm{b}}}}_k=\sqrt{P_k}\frac{(\bar{\boldsymbol{\mathrm{\Lambda}}}_{k}^{\text{ZF}})^{-1}\tilde{{\boldsymbol{\mathrm{H}}}}_{k,\rho}^H[0]\bar{{\bf{w}}}_k}{\|(\bar{\boldsymbol{\mathrm{\Lambda}}}_{k}^{\text{ZF}})^{-1}\tilde{{\boldsymbol{\mathrm{H}}}}_{k,\rho}^H[0]\bar{{\bf{w}}}_k\|}. \label{rd111}
\end{align}

Based on the obtained results, the overall MMSE-based beamforming for solving problem \eqref{lblvvv} is summarized in Algorithm \ref{sca}. Note that in each iteration, the variables are optimally obtained with
other variables being fixed, and the resulting values of problem \eqref{lblvvv} are non-decreasing, an thus Algorithm \ref{sca} is guaranteed to converge.

\begin{algorithm}[t]
	\caption{Alternating optimization for ISI-ZF path-based transmit beamforming and receive combining}
	\label{sca}
	\begin{algorithmic}[1]
		\State Initialize a feasible solution $\|\bar{{\bf{b}}}_k^{(r)}\|^2=P/K, \forall k$. Let $r=0$.
		\Repeat
		\State {Obtain the receive combining vector ${\boldsymbol{\mathrm{w}}}_k^{(r+1)}, \forall k$ based on \Statex\hspace{5 mm}  \eqref{rd11}. }
		\State {Obtain the transmit beamforming vector $\bar{\boldsymbol{\mathrm{b}}}_k^{(r+1)}, \forall k$ based   \Statex\hspace{5 mm} on \eqref{rd111}. }
		\State { Set $r=r+1$}.
		\Until{ The  fractional increase of objective value of problem \eqref{lblvvv}  is below a certain threshold}.
	\end{algorithmic}
\end{algorithm}

In the following, the eigen-beamforming and ISI-ZF beamforming designs for the benchmarking scheme is  considered.

 \begin{figure*}[h]
	\centering
	
	\subfigure[$M_t=128$ and $M_r=2$]{
		\begin{minipage}[t]{0.49\linewidth}
			\centering
			\includegraphics[scale=0.06]{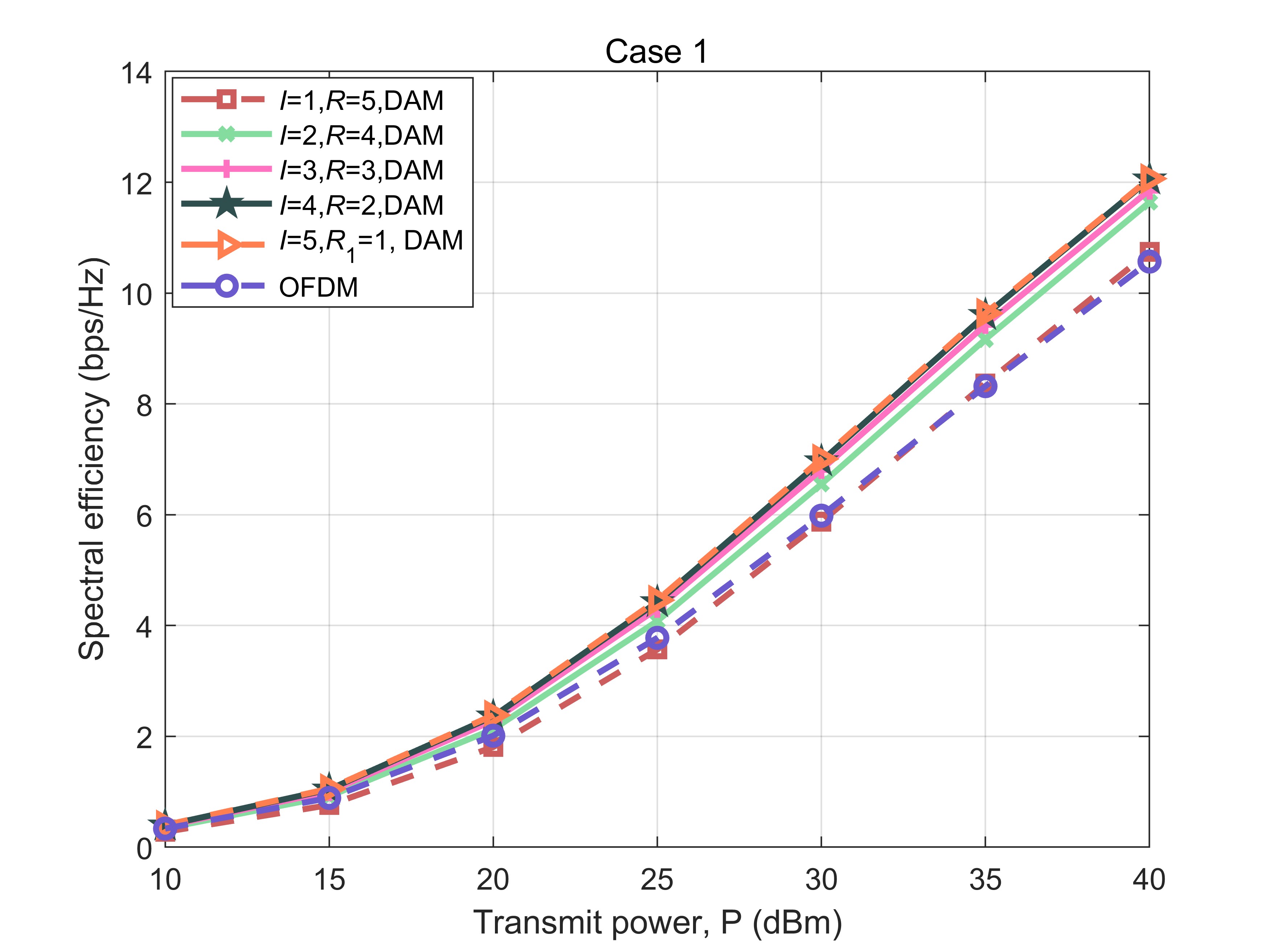}
		\end{minipage}%
	}%
	\subfigure[$M_t=4$ and $M_r=64$]{
		\begin{minipage}[t]{0.49\linewidth}
			\centering
			\includegraphics[scale=0.06]{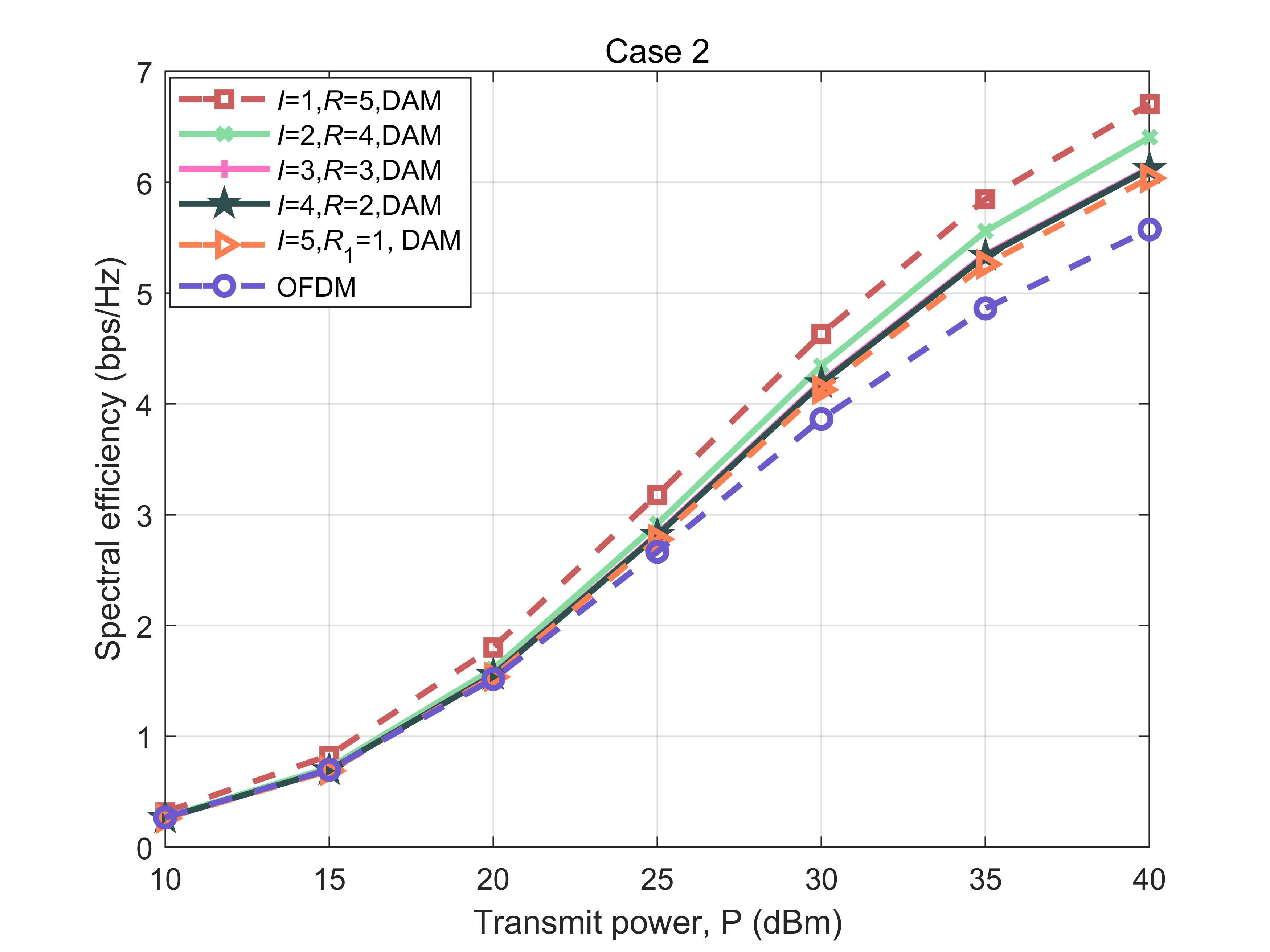}
		\end{minipage}%
	}%
	
	\subfigure[$M_t=128$ and $M_r=64$]{
		\begin{minipage}[t]{0.49\linewidth}
			\centering
			\includegraphics[scale=0.06]{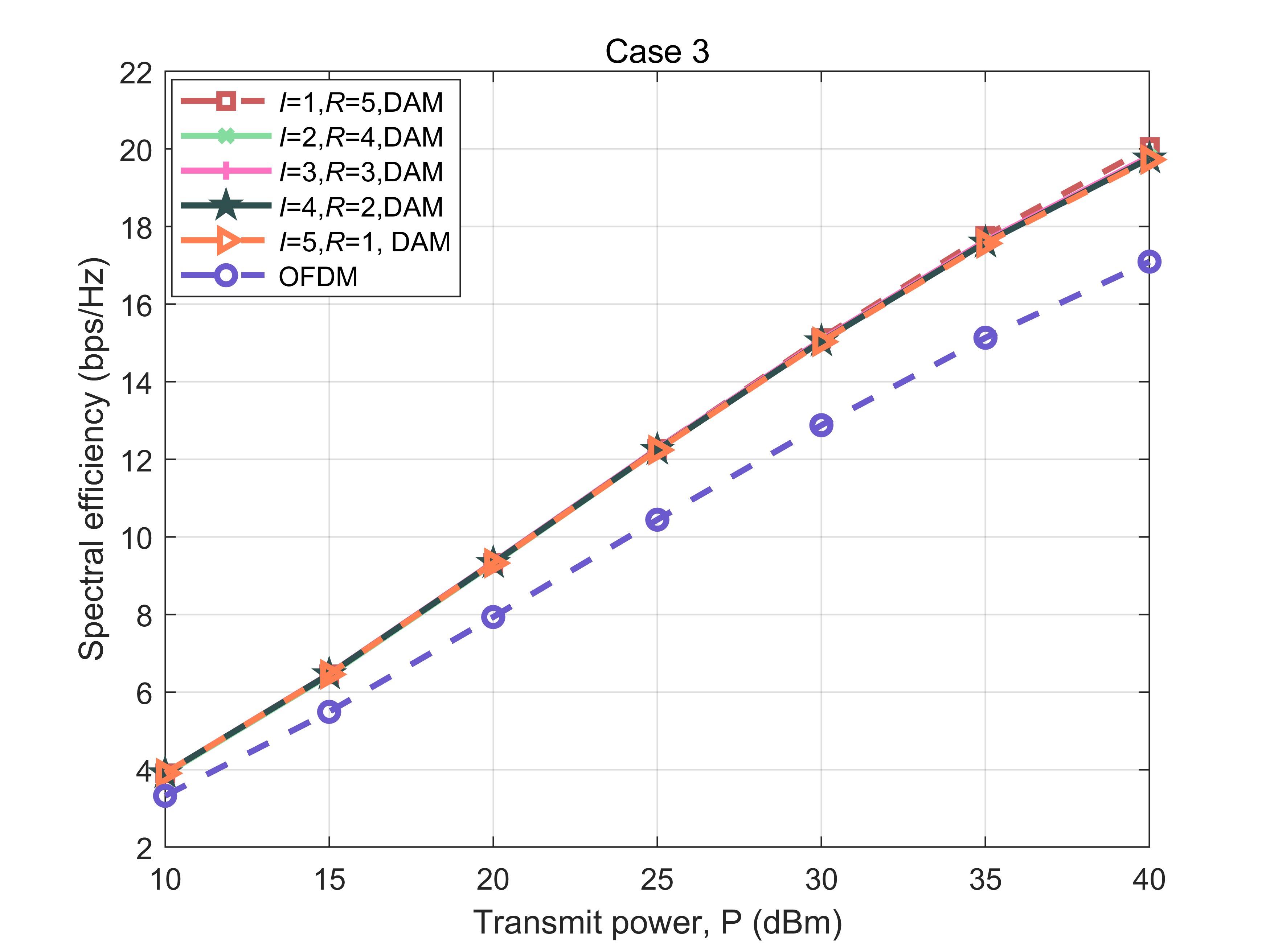}
		\end{minipage}
	}%
	\subfigure[$M_t=4$ and $M_r=2$]{
		\begin{minipage}[t]{0.49\linewidth}
			\centering
			\includegraphics[scale=0.06]{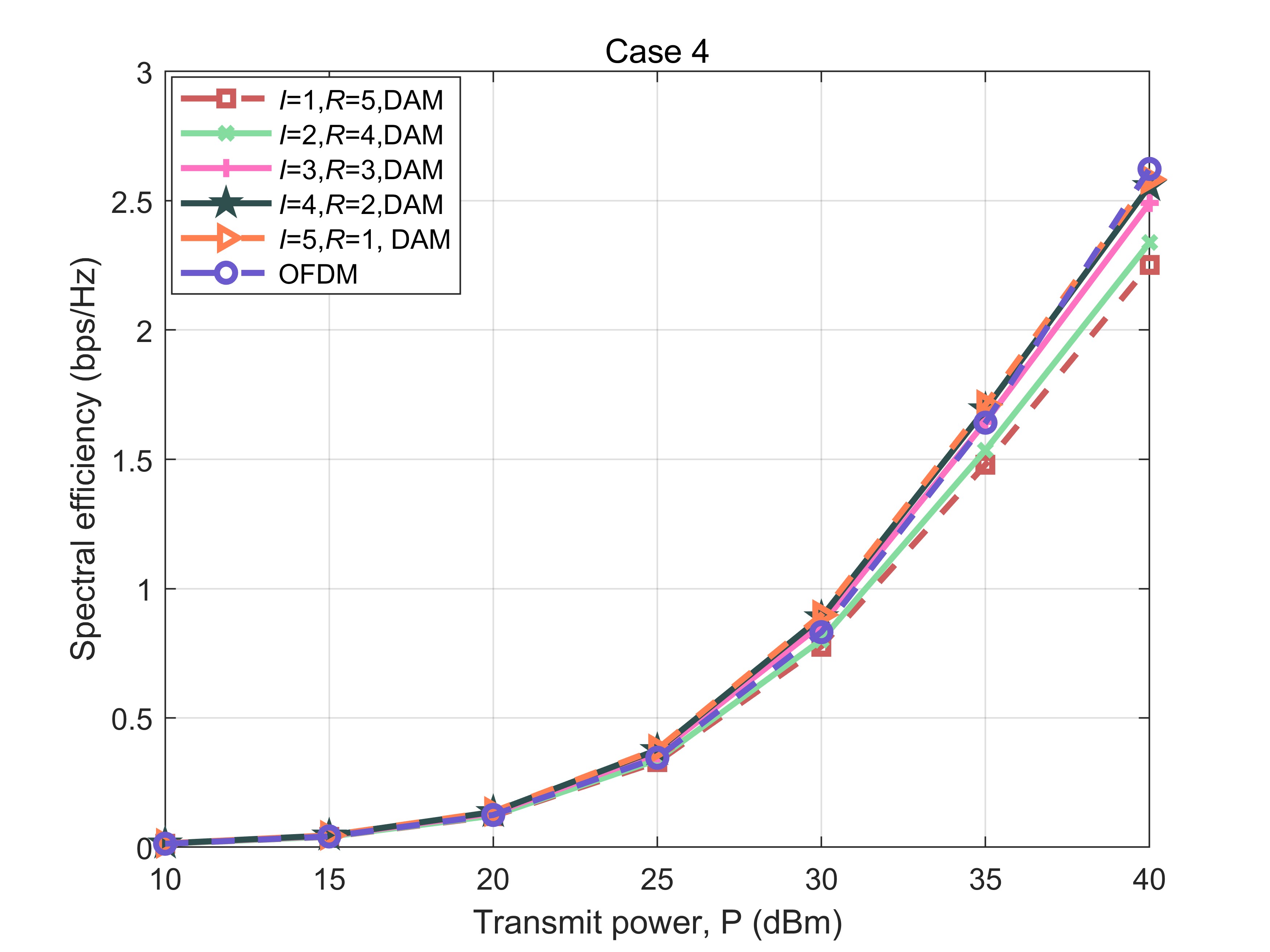}
		\end{minipage}
	}%
	
	\centering
	\caption{Spectral efficiency versus transmit power  for  the proposed double-side multi-user DAM and OFDM.} \label{fig4}
\end{figure*}

 \subsection{Bemchmarking scheme: multi-user OFDM} \label{fffn}
 \subsubsection{Eigen-beamforming transmission}\label{fff2}
 For the benchmarking OFDM scheme, let $M$ denote the number of sub-carriers,  and $s_{k}[d,m]$ denote the information-bearing symbol of UE $k$ at  the $m$-th sub-carrier of the $d$-th OFDM symbol, $d=-\infty,..,+\infty, m=0,...,M-1$, with $\mathbb{E}[|s_{k}[d,m]|^2]=1$.
 The transmitted signal of the BS  at $m$-th sub-carrier for the $d$-th OFDM symbol  is
 \begin{align}
 	&{\boldsymbol{\mathrm{s}}}[d,m]=\sum_{k=1}^K\boldsymbol{\mathrm{v}}_{k,m}s_{k}[d,m], \forall d,m, \\ \notag
 \end{align}
 where  $\boldsymbol{\mathrm{v}}_{k,m} \in \mathbb{C}^{M_t \times 1}$ denotes the frequency-domain beamforming vector  for UE $k$ at the $m$-th sub-carrier. The corresponding  power constraint  is
 \begin{equation}
 	\sum_{m=0}^{M-1}\mathbb{E}[\|{\boldsymbol{\mathrm{s}}}[d,m]\|^2]=\sum_{m=0}^{M-1}\sum_{k=1}^{K}\|\boldsymbol{\mathrm{v}}_{k,m}\|^2\leq MP \label{RROFDM-3}.
 \end{equation}
 
 By applying \( M \)-point discrete Fourier transform (DFT) of \eqref{1}, the  frequency domain channel is
\begin{align}
	\boldsymbol{\mathrm{H}}_{k,m}=\frac{1}{\sqrt{M}}\sum_{l=1}^{L_k} \boldsymbol{\mathrm{H}}_{kl}e^{j2\pi \frac{mn_{kl}}{M}}.
\end{align}

After the receive combining $\boldsymbol{\mathrm{u}}_{k,m}$,  the resulting frequency domain received signal is 
  \begin{align} 
 	&	y_k[d,m]=\boldsymbol{\mathrm{u}}_{k,m}^H\boldsymbol{\mathrm{H}}_{k,m}\sum_{k'=1}^{K}\boldsymbol{\mathrm{v}}_{k',m} s_{k'}[d,m] + {z}_k[d,m],\label{vvvv1}  \\ \notag
 \end{align}
 where $z_k[d,m]\sim \mathcal{CN}(0,\hat{\sigma}^2)$ is the AWGN , with  $\hat{\sigma}^2={\sigma^2}/{M}$.

 Hence, the resulting SINR is
 \begin{align}
 	&\gamma_{k,m}=\frac{|\boldsymbol{\mathrm{u}}_{k,m}^H\boldsymbol{\mathrm{H}}_{k,m}\boldsymbol{\mathrm{v}}_{k,m}|^2}{\sum_{k'\ne k}^{K}|\boldsymbol{\mathrm{u}}_{k,m}^H\boldsymbol{\mathrm{H}}_{k,m}\boldsymbol{\mathrm{v}}_{k',m}|^2+\hat{\sigma}^2\|\boldsymbol{\mathrm{u}}_{k,m}\|^2}. \label{006}
 \end{align}

Similarly, for $\boldsymbol{\mathrm{H}}_{k,m}= {{\bf{U}}}_{k,m}{{\bf{\Sigma}}}_{k,m}{{\bf{V}}}_{k,m}^H$,   the transmit beamforming and receive combining vectors for the eigen-beamforming transmission are ${{\boldsymbol{\mathrm{v}}}}_{k,m}=\sqrt{MP}[{{\bf{V}}}_{k,m}]_{:,1}/\|{\bf{V}}_{f,m}\|_{\text{F}}$ with ${\bf{V}}_{f,m}=\left[[{{\bf{V}}}_{1,m}]_{:,1},...,[{{\bf{V}}}_{K,m}]_{:,1}\right]$, and ${{\boldsymbol{\mathrm{u}}}}_{k,m}=\left[{{\bf{U}}}_{k,m}\right]_{:,1}/\left\|[{{\bf{U}}}_{k,m}]_{:,1}\right\|$, respectively, and then the resulting sum rate  can be  obtained.

\subsubsection{ZF beamforming}\label{ff1}

The sub-carrier based transmit beamforming is similarly denoted as ${\boldsymbol{\mathrm{v}}}_{k,m}=\bar{{\boldsymbol{\mathrm{H}}}}_{k,m}^{\perp}{\bf{b}}_{k,m}, \forall k,m$, where $\bar{{\boldsymbol{\mathrm{H}}}}_{k,m}^{\perp}\in \mathbb{C}^{M_t\times N_t}$ denotes an orthogonal basis for the orthogonal complement of  $\bar{{\boldsymbol{\mathrm{H}}}}_{k,m}$, with $\bar{{\boldsymbol{\mathrm{H}}}}_{k,m}=[{\boldsymbol{\mathrm{H}}}_{1,m}^H,...,{\boldsymbol{\mathrm{H}}}_{k-1,m}^H,{\boldsymbol{\mathrm{H}}}_{k+1,m}^H,...,{\boldsymbol{\mathrm{H}}}_{K,m}^H]\in \mathbb{C}^{M_t \times (K-1)M_r}$, $N_t=\text{rank}(\bar{{\boldsymbol{\mathrm{H}}}}_{k,m}^{\perp})$, and ${\bf{b}}_{k,m}\in \mathbb{C}^{N_t\times 1}$   is the transmit beamforming vector to be optimized. The feasibility of the ZF constraint is ensured when $M_t\geq (K-1)M_r+1$. Substituting ${\boldsymbol{\mathrm{v}}}_{k,m}=\bar{{\boldsymbol{\mathrm{H}}}}_{k,m}^{\perp}{\bf{b}}_{k,m}$ into \eqref{vvvv1}, we have
\begin{align}
	&y_k[d,m] =\boldsymbol{\mathrm{u}}_{k,m}^H\boldsymbol{\mathrm{H}}_{k,m}\bar{{\boldsymbol{\mathrm{H}}}}_{k,m}^{\perp}{\bf{b}}_{k,m}s_{k}[d,m] + {z}_k[d,m].  \label{881v} \\ \notag
\end{align}
Consequently, the signal-to-noise ratio (SNR) of  sub-carrier $m$ for UE $k$ is 
\begin{align}
	\gamma_{k,m}^{\text{ZF}}=\frac{|\boldsymbol{\mathrm{u}}_{k,m}^H\boldsymbol{\mathrm{H}}_{k,m}\bar{{\boldsymbol{\mathrm{H}}}}_{k,m}^{\perp}{\bf{b}}_{k,m}|^2}{\sigma^2\|{\boldsymbol{\mathrm{u}}}_{k,m}\|^2 }.
\end{align}
The sum rate can be maximized by optimizing the transmit beamforming vector ${{\bf{b}}}_{k,m}, \forall k, m$ and receive combining vector ${{\boldsymbol{\mathrm{u}}}}_{k,m}, \forall k, m$, which yields the following problem
\begin{align}
	\max\limits_{{{\bf{u}}}_{k,m}, {{\bf{b}}}_{k,m} \forall k} ~&\frac{1}{M}\sum_{m=0}^{M-1}\sum_{k=1}^K\text{log}_2\left(1+\frac{|\boldsymbol{\mathrm{u}}_{k,m}^H\boldsymbol{\mathrm{H}}_{k,m}\bar{{\boldsymbol{\mathrm{H}}}}_{k,m}^{\perp}{\bf{b}}_{k,m}|^2}{\sigma^2\|{\boldsymbol{\mathrm{u}}}_{k,m}\|^2 }\right)\label{lbl2} \\ \notag
	\text{s.t.}&\begin{matrix}\sum_{m=0}^{M-1}\end{matrix}\begin{matrix} \sum_{k=1}^K\end{matrix}\|{\boldsymbol{\mathrm{b}}}_{k,m}\|\leq MP\tag{\ref{lbl2}{a}} \label{lbl2a},\\ \notag
\end{align}
which can be effectively  solved by SVD and the water-filling (WF) algorithm.

\subsection{Guard interval}
Denote by $G_c= T_c/T$ the number of signal samples  within each channel coherence time,  and $T_c$ represents the channel coherence time.  Then, a CP of length $G_{\text{CP}}$ is required for each OFDM symbol \cite{MU-DAM}. Thus, each OFDM symbol duration is $(M+G_{\text{CP}})T$, and the number of OFDM symbols in each channel coherence time is $n_{\text{OFDM}}=G_c/(M+G_{\text{CP}})$.  The guard interval overhead of OFDM $n_{\text{OFDM}}G_{\text{CP}}/G_c=G_{\text{CP}}/(G_{\text{CP}}+M)$.
The effective spectral efficiency   is then given by
\begin{align}
	R_{\text{OFDM}} &=\frac{G_c-n_{\text{OFDM}}G_{\text{CP}}}{G_c}\sum_{k=1}^K\sum_{m=1}^M \text{log}_2(1+\gamma_{k,m}) \label{GI-1},
\end{align}
where   $\gamma_{k,m}$ denotes the SINR/SNR of UE $k$ at the  $m$-th sub-carrier for OFDM transmission.	

 To avoid the inter-block interference, the single-carrier   multi-user DAM  only need a guard interval $G_{\text{GI}}$ at the beginning  of each channel coherence block \cite{DAM-OFDM}. By considering the guard interval overhead, the effective spectral efficiency of single-carrier transmission  is
\begin{equation}
	R_{\text{DAM}} = \frac{1}{1+\beta}\frac{G_c- G_{\text{GI}}}{G_c}\sum_{k=1}^K\text{log}_2(1+\gamma_k), \label{GI-2}
\end{equation} 
where $\beta$ denotes the roll-off factor, and  $\gamma_k$ is the SINR of UE $k$ for multi-user DAM.

\section{Simulation results}\label{simulation} 

In this section, we present simulation results to evaluate the performance of the proposed multi-user double-side DAM scheme. We operate within a system at a frequency of $f_c=$ 28 GHz, with a signal sampling  interval of $T=5$ ns. Utilizing a raised cosine pulse filter characterized by a roll-off factor of $\beta = 0.01$, we achieve a total bandwidth of $B=(1+\beta)f_B=$ 210 MHz, with $f_B=\frac{1}{T}=$200MHz. The noise power spectral density is denoted as $N_0=-174$ dBm/Hz, yielding a total noise power of $\sigma^2=-91$ dBm. Our channel coherence time spans $T_c = 1$ ms, accommodating a total of $G_c = 2 \times 10^5$ signal samples. Unless specified otherwise, the transmit power at the BS is set to $P=30$ dBm, with $K =2$ UEs. Each UE experiences $L_k=L=3$ temporal-resolvable multi-paths, with their delays randomly distributed within the range $[0, 100T]$, and thus $G_{\text{CP}}=100$. Note that  the overhead introduced by the guard interval
between different channel coherence blocks is $({G_c- G_{\text{GI}}})/{G_c}\to 1$, with $ G_{\text{GI}}=200$ \cite{DAM}, and thus the overhead is determined by roll-off factor \eqref{GI-2}. 
Additionally, the angles of departure (AoDs) for all multi-paths are randomly selected from $[-90^\circ,90^\circ]$, and the complex-valued gains for each path are generated based on the model elaborated in \cite{ChanneModel}. The number of sub-carriers for OFDM  is  $M = 512$.  Besides, in Section \ref{fffn}, the use of integer channel delays and optimal power allocation establishes OFDM as the upper performance bound for scenarios involving fractional channel delays, hence it is referred to as the ``upper bound OFDM" .

\begin{figure}[htp]
	\centering
	\includegraphics[scale=0.07]{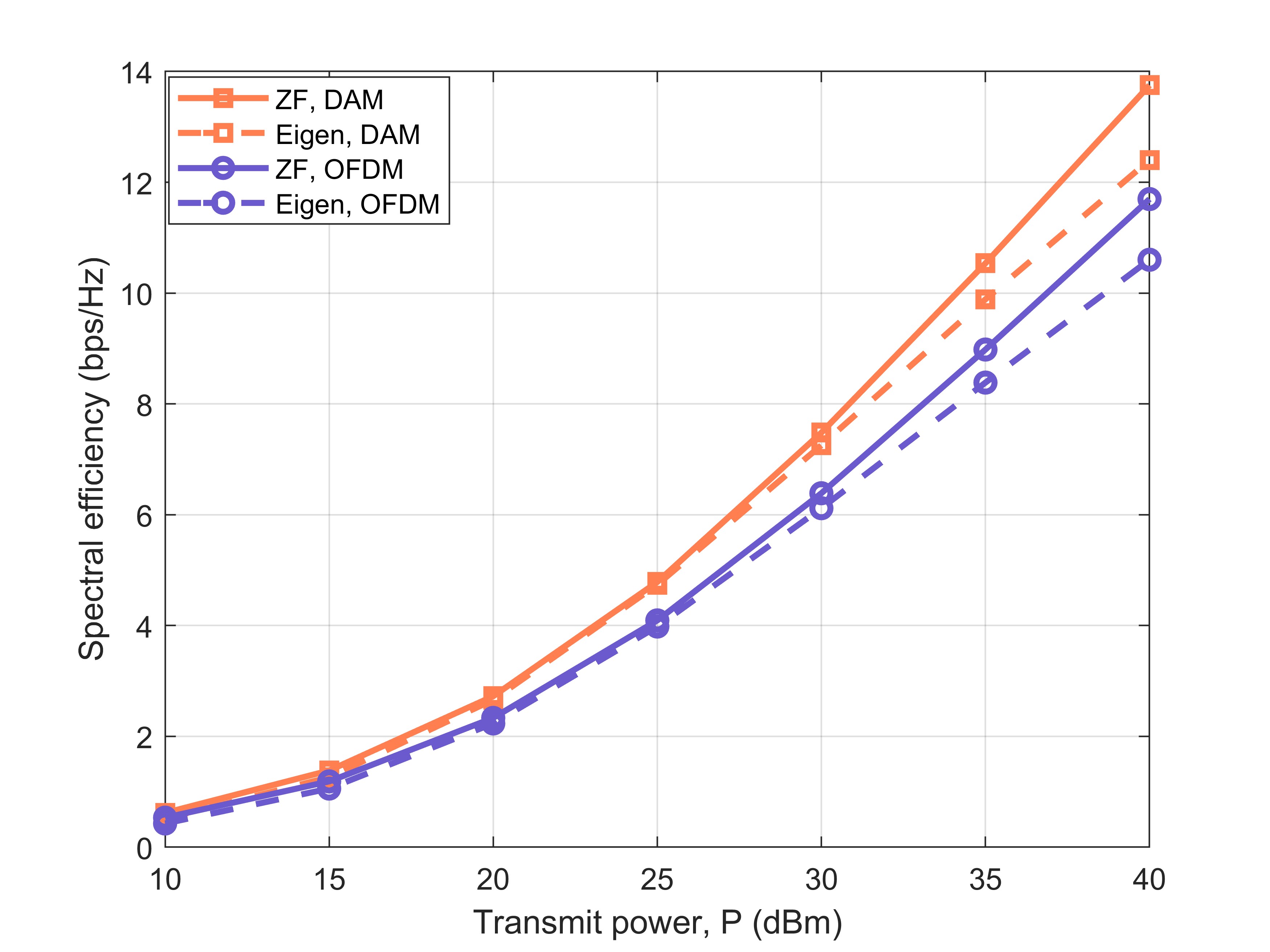}
	\caption{Spectral efficiency versus transmit power for the proposed BS-side multi-user DAM and OFDM with the integer channel delays.} \label{fig6}
\end{figure}

\begin{figure}[htp]
	\centering
	\includegraphics[scale=0.07]{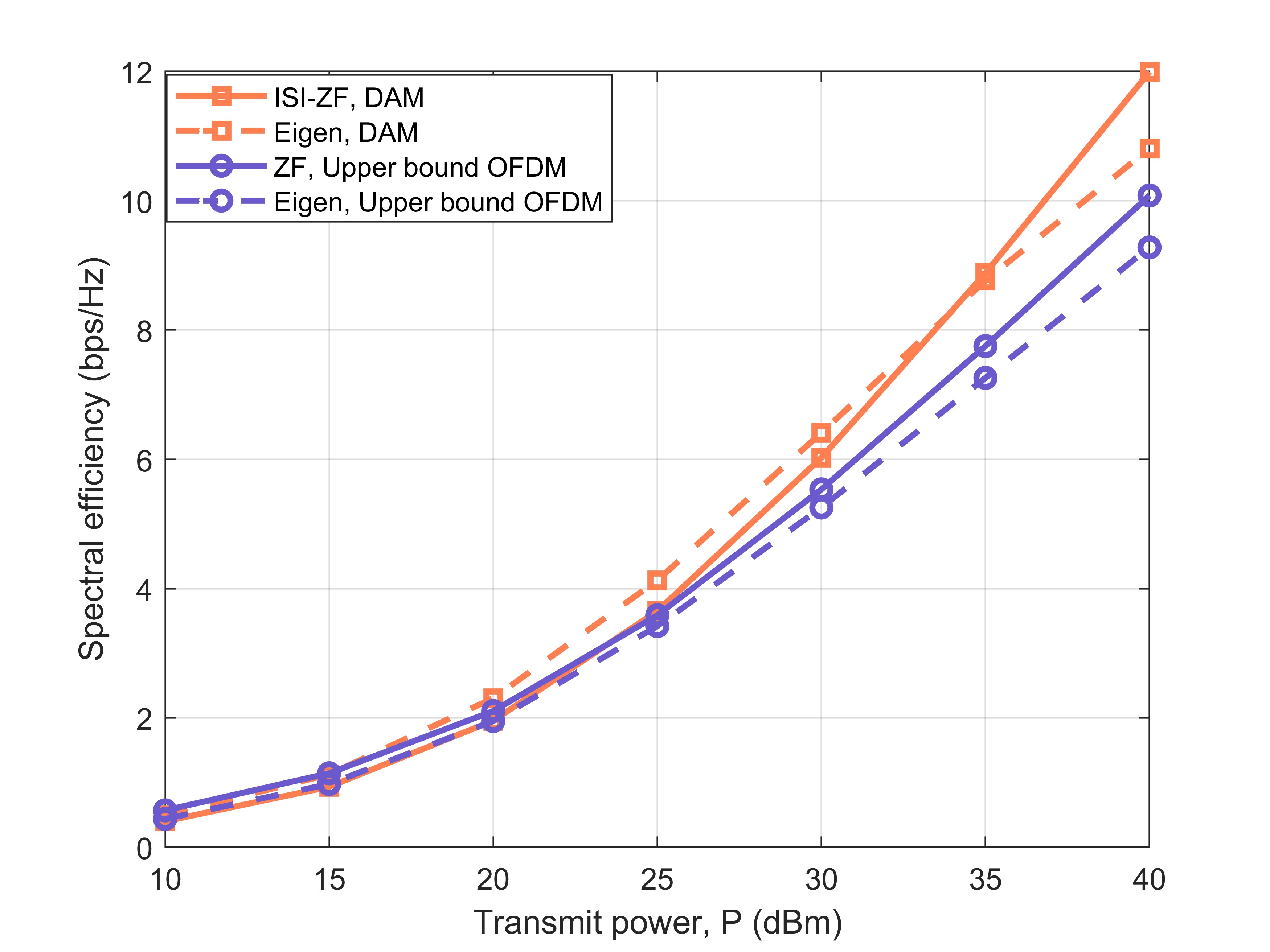}
	\caption{Spectral efficiency versus the transmit power for the proposed BS-side multi-user DAM with the fractional channel delays and the upper bound OFDM .}\label{fig7}
\end{figure}

Fig. \ref{fig4} illustrates the spectral efficiency versus the transmit power for the proposed BS-side DAM, UE-side DAM and double-side DAM in Section \ref{rrrr1}, together with the benchmarking scheme of  OFDM. In Fig. \ref{fig4}(a), where $M_t=128$ and $M_r=2$, the introduction of additional delay components at the BS side significantly enhances the performance of multi-user DAM relative to OFDM. Specifically, the more delays introduced, the better the performance is. This improvement is attributed to the large number of antennas at the BS, which can more effectively resolve additional delay components compared to the limited number of antennas at the UE side. Similarly, in Fig. \ref{fig4}(b), when the UE is equipped with a larger number of antennas ($M_t=4$ and $M_r=64$), UE-side DAM achieves better performance compared to double-side DAM, BS-side DAM, and OFDM. In scenarios where both the BS and UE sides are equipped with substantial antenna arrays ($M_t=128$ and $M_r=64$), as shown in Fig. \ref{fig4}(c), BS-side DAM, UE-side DAM, and double-side DAM all exhibit excellent performance, consistently outperforming OFDM. These results align with those derived from Case 1, Case 2, and Case 3. Furthermore, as depicted in Fig. \ref{fig4}(d), when both the BS and UE have a limited number of antennas ($M_t=4$ and $M_r=2$), distributing the delay components across both sides ($I_1 = I_2 = 4$, Case 4) is shown to be advantageous for enhancing performance. Nonetheless, BS-side DAM continues to demonstrate strong performance in this configuration, owing to the interference minimization model for problem \eqref{122}, which neglects IUI.

Fig. \ref{fig6} depicts the spectral efficiency versus the  transmit power for the proposed multi-user DAM and OFDM, considering integer channel delays. As shown in Fig. \ref{fig6}, with $M_t=128$ and $M_r=2$, the performance of  the multi-user DAM scheme outperforms that of OFDM. This performance gain is primarily due to the efficient utilization of all available multi-path components.

Fig. \ref{fig7}  presents the spectral efficiency versus the  transmit power for the proposed multi-user DAM with fractional channel delays, and the upper bound  OFDM with $M=256$. As observed in Fig. \ref{fig7}, when channel delays become fractional, the performance of the ISI-ZF multi-user DAM is degraded due to the introduction of additional ISI from integer delay pre-compensation,  the local convergence in the alternating optimization process, and the use of sub-optimal equal power allocation among UEs compare with Fig. \ref{fig6}. In contrast, OFDM is unaffected by fractional time delays and benefits from WF power allocation. However, the performance of the proposed ISI-ZF beamforming and eigen-beamforming schemes of multi-user DAM remains superior to that of OFDM with $M = 256$. 
Although compensating for fractional channel delays with integer multiples introduces some tailing for multi-user DAM, the approach still benefits from the effective utilization of all multi-path components. Compared to Fig. \ref{fig6}, the performance of OFDM declines as the number of sub-carriers decreases. Furthermore, this performance degradation continues with a further reduction in sub-carriers, whereas multi-user DAM remains unaffected by the number of sub-carriers.

\begin{figure}[htp]
	\centering
	\includegraphics[scale=0.07]{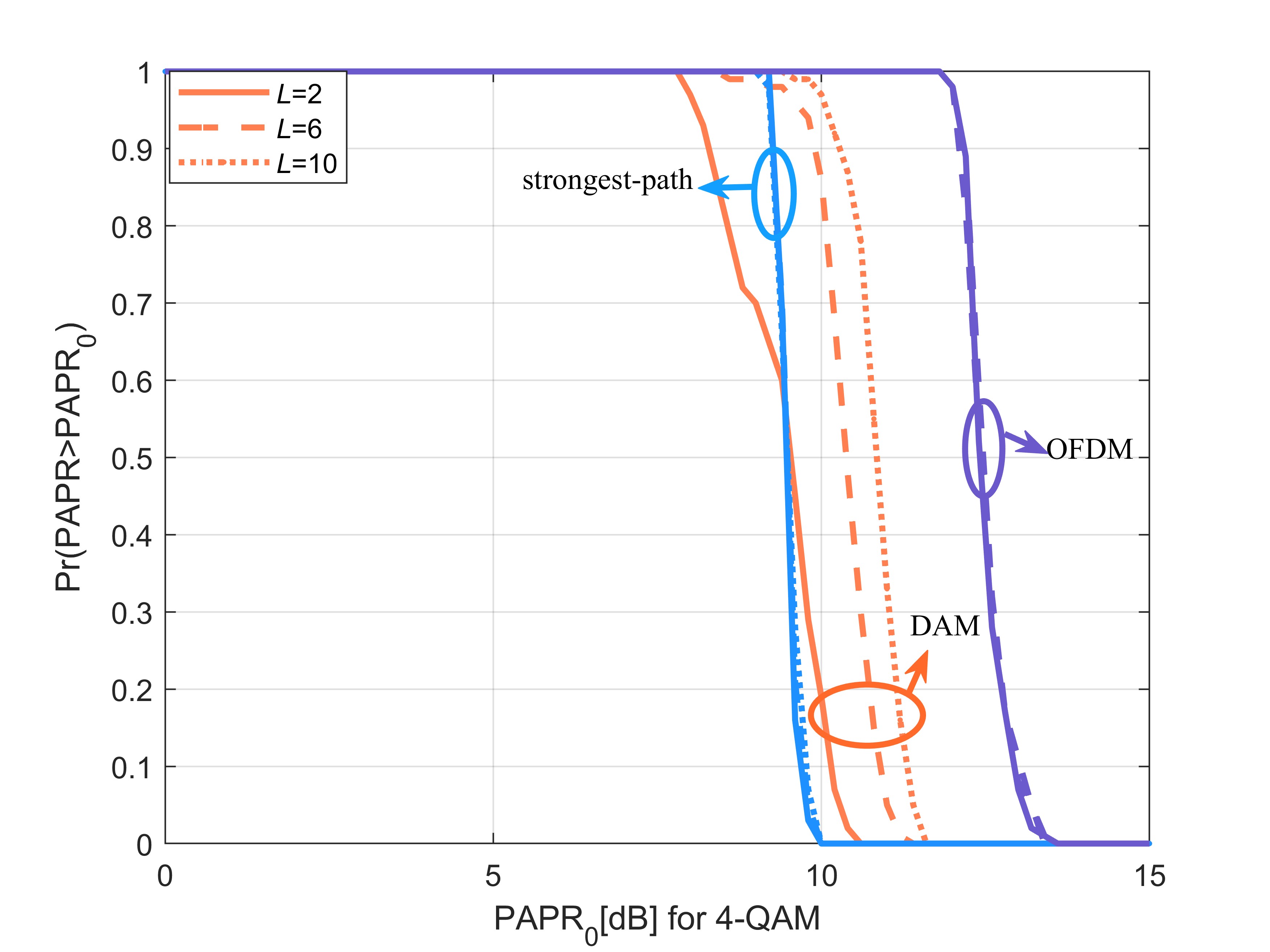}
	\caption{PAPR comparison for BS-side multi-user DAM,  OFDM, and the strongest path based scheme with 4-QAM modulation, where $M_t=128, M_r=2$.}\label{fig8}
\end{figure}

Fig. \ref{fig8}  presents a comparison of the PAPR for BS-side multi-user DAM,  OFDM with $M=512$, and the strongest path based scheme \cite{MU-DAM} employing corresponding eigen-beamforming transmission strategies, all utilizing 4 quadrature amplitude modulation (QAM). Besides, the time domain signals for the OFDM and strongest path based scheme  adopt identical transmit pulse shaping filtration as employed in multi-user DAM. Notably, the PAPR is calculated in the analog domain after transmit filtering. As depicted in Fig. \ref{fig8}, the multi-user DAM technique exhibits a significantly lower PAPR compared to OFDM. This outcome is anticipated since, in OFDM, a total of $KM$ signals are superimposed on each antenna, whereas in DAM, ${L}_{\mathrm{tot}}$ multi-path signals are combined on each antenna.

\section{Conclusion}

This paper investigated the performance of multi-user DAM with both delay pre-compensation and post-compensation in the presence of fractional channel delays in mmWave/THz massive MIMO systems. Initially, we established the relationship between the number of pre-compensation and post-compensation  and that of the multi-path channel components. For a given number of delay pre/post-compensation, we determined the corresponding pre-compensation and post-compensation vectors. Our analysis showed that when the number of BS/UE antennas is sufficiently large, single-side DAM—where delay compensation is only performed at the BS/UE—is preferable to double-side DAM, as it results in less ISI requiring spatial elimination. We also proposed two low-complexity path-based beamforming strategies based on eigen-beamforming transmission and ISI-ZF principles, respectively, and evaluated the achievable sum rate. Simulation results demonstrated that with sufficiently large BS/UE antennas, single-side DAM is sufficient. Moreover, BS-side multi-user DAM exhibits higher spectral efficiency and/or lower PAPR compared to OFDM.

\bibliographystyle{elsarticle-num}
\balance
\bibliography{Reference}

\end{document}